\documentclass[12pt]{article}

\usepackage{a4}
\usepackage{graphics}
\usepackage{amsfonts,amssymb}

\newcommand{\half}{\mbox{\scriptsize$1\over2$}}
\newcommand{\imi}{\mathrm{i}}
\newcommand{\dex}{\mathbin{\mathrm{d}}}

\newcommand{\Fn}{{\cal F}}
\newcommand{\halfi}{{\textstyle{\mathrm{i}\over2}}}

\newcommand{\zeroth}{{(0)}}
\newcommand{\idc}{\iota_{\mbox{\scriptsize dc}}}

\newcommand{{\chao}}{`irregular'}
\newcommand{\phaselocking}{phase-locking}
\newcommand{\ie}{{\it i.e.}\/}

\begin{document}

\begin{quote}
\begin{center}\large
Long-term behavior \\
of solutions of the
equation $\dot \phi + \sin\phi=f$ \\
with periodic $f$ \\
and the modeling of dynamics \\
 of
overdamped Josephson junctions:\\
\it
 Unlectured notes
\end{center}
\end{quote}

\begin{center}
S.I. Tertychniy\\
Russia, VNIIFTRI
\end{center}

\begin{abstract}\noindent
The 
method of efficient description of long-term
behavior of solutions of the non-linear
first order ODE $\dot \phi + \sin\phi=f$
for arbitrary periodic  $f$
is discussed.
The criterion enabling one to separate and identify the qualitatively
different
solutions
is established.
The applications of the method to the modeling of
dynamics
of overdamped Josephson junctions
in superconductors
are outlined.
\end{abstract}

\section*{Preliminaries}

A simple model of a Josephson junction \cite{X4}
proposed by Stewart and  McCumber \cite{X2,X3}
is based, from the mathematical point of view, on
the non-linear second order ODE
 \begin{equation}
 \beta\ddot\phi+\dot \phi + \sin\phi=f.
                                  \label{eq1}
\end{equation}
Here $\phi=\phi(t)$ is the unknown function (called
{\it phase}) representing the
difference of phases of the
so called order parameter which plays the role of the macroscopic wave function
describing the states of two weakly coupled superconducting electrodes
constituting  Josephson junction.
The connection of the phase with observable quantities
is established by the
 Josephson equation \cite{X1} which connects the
voltage across the junction with the
time derivative of $\phi$:
 \begin{equation}
V={\Phi_0\over 2\pi}{d\phi\over d \tau}\propto\dot\phi\equiv{d\phi\over d t}.
                                  \label{eq2}
\end{equation}
Here the physical constant $\Phi_0=h(2e)^{-1}$ ($h$ is the Plank constant, $e$ is the
electron charge) is
called the (magnetic) flux quantum,
the parameter $t$ used in  Eq.\ (\ref{eq1}) is
the
rescaled dimensionless time 
 $t=\omega_c \tau$,
 where $\tau$ is the `genuine' (dimensional) time,
the
constant $\omega_c$ 
called characteristic
frequency is connected with junction properties. It can be defined
by the equation $\omega_c=2\pi\Phi_0^{-1}R_{N}I_c$,
where $R_N,I_c$ are the junction characteristics, namely,
$R_N$ is its resistance in the normal
 state,
$I_c$ is called critical
current (it estimates the maximal current which can flow through 
the junction without
application of external voltage).
The positive constant
$\beta$
is
connected with the junction capacitance, the less the latter, the less
the former.
The function $f=f(t)$ 
represents the current
(normalized to $I_c$ and named {\em bias})
circulating in
the junction circuit and
supplied by an external source.
It is assumed to be specified in advance.

Here we consider the case of a periodic bias
of the known period $T$ and arbitrary profile.
Thus  $f$ is assumed to satisfy the condition
 \begin{equation}
f(t)=f(t+T).
\end{equation}
Usually $f$ is a continuous (or smooth) 
function but a finite number of finite jumps
on the period interval
is also allowed\footnote{
Having added of a finite number of Dirac $\delta$-like pulses,
the problem retains
meaningful 
but such a generalization will not be pursued here.
}.

It is worth noting that
there is a common practice
in the physical literature
to divide
the periodic bias function $f$ into some
constant constituent
(`direct current', DC) $\iota_{\mbox{\scriptsize dc}}$
and the
residual  one
$\iota_{\mbox{\scriptsize ac}}$ (alternating, high frequency, rf current, AC):
 \begin{equation}
f=\iota_{\mbox{\scriptsize dc}}+\iota_{\mbox{\scriptsize ac}}.
\end{equation}
Following this convention,
we assume for definiteness that
the average value of $f$ is assigned to
$\iota_{\mbox{\scriptsize dc}}$.
Accordingly,  by definition, the average value
of the residual
$\iota_{\mbox{\scriptsize ac}}$ vanishes%
 \footnote{Such an interpretation  reflects
the typical experimental setup which fixes
$\iota_{\mbox{\scriptsize ac}}$
and considers $\iota_{\mbox{\scriptsize dc}}$
as a free parameter to be varied.
In particular,
the current-voltage (I-V) curve
of a Josephson junction
is understood 
just as the
dependence of the average voltage across it 
on the DC bias contribution $\iota_{\mbox{\scriptsize dc}}$ whereas 
$\iota_{\mbox{\scriptsize ac}}$ is kept unchanged.}.

The important property of a Josephson junction, which
is perfectly
captured by the Stewart-McCumber model in spite of its
comparative
homeliness,
is the {\em \phaselocking{}\/} effect.
Qualitatively, the  \phaselocking{}
may be regarded as
the relaxation 
in the course of the phase evolution
of the `loose
constituent' of the phase function
reflecting its initial state.
Then
the only surviving dynamical `process' 
proves
a steady one  `dragged' by the periodic
bias and not depending of the past phase evolution.
Said
another way,
once some lapse depending on
the specific rate of
relaxation of 
initial perturbations 
and the `magnitude' of the latter
elapses, the phase
function profile proceeds with the
{\it reproducing itself\/}, modulo a $2\pi$-aliquot contribution,
on each subsequent time step of duration $T$.
Obviously, such a situation would take place,
in particular,
if,
regardless
of the initial state
the phase starts to evolve from,
$\phi(t)$ 
converges 
 for large $t$
to a  periodic function
of the period $T$.
There are also other, non-periodic (but still fairly specific as we shall see
below)
forms of 
such an asymptotically `steady' phase evolution.
On the other hand, 
the \phaselocking{} property
is in no way a universal one
to be automatically attributed
to solutions of Eq.\ (\ref{eq1}).
Depending on the model parameters, the phase evolution
may be completely different,
revealing, in particular, no periodicity or any other 
apparent order
\cite{X5}.

\begin{figure}
\includegraphics{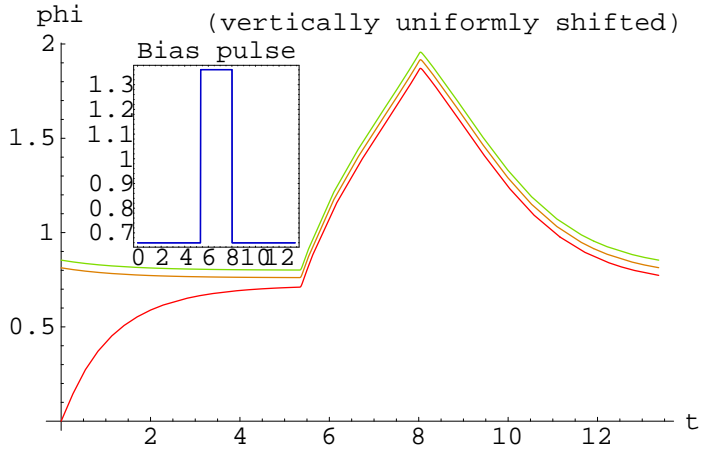}
\caption{\small
Three subsequent segments (red, brown=$\half$(red+green), green)
of the single graph describing evolution
of the phase function on the time interval $[0,3T]$ are
placed over the common $t$-axes segment
$[0,T]$
by means of appropriate `horizontal' shifts `backwards in time'
({\it plot segmenting}).
The specific parameter value is
$\iota_{\mbox{\scriptsize dc}}=0.8$
the values of other parameters
are given in the main text.
In order to visually resolve the neighbor graphs
here (and in  subsequent similar plots)
all the curves, except for a starting one,
are uniformly shifted
in the `vertical' direction
by an additional amount of space 
which is uniformly incremented for each subsequent graph curve
(here by 0.04 units).
As one can see,
the third (green) graph segment already coincides
with the second one, essentially, yielding
therefore the (approximate) specimen of the
limiting phase evolution function which
proves here periodic.
}
\label{f1}
\end{figure}

 The plots shown in
figure \ref{f1}
illustrates the simplest case of the \phaselocking{} in a numerical example.
Here the result
of a straightforward
numerical integration of Eq.\ (\ref{eq1})
is displayed. For the sake of definiteness,
the bias function $f(t)$ is chosen to
be the periodic rectangular pulse sequence,
one of the pulses being shown in the inset.
Its integral magnitude amounts to 1.5, the frequency is
$2\pi/T=0.47$ ($T\simeq13.3685$), $\beta=0.02$, the width of the
pulse peak
(here being rather a plateau
symmetric with respect to the plot center)
amounts to 20\%{} of the total
pulse period. The `constant bias constituent'
$\iota_{\mbox{\scriptsize dc}}$ (the average of $f$)
equals 
$0.8$.

The phase evolution displayed in Fig.\ \ref{f1}
starts with the null initial conditions
$\phi(0)=0=\dot\phi(0)$ chosen
for
simplicity reasons%
 \footnote{It is worth noting that the condition $\dot\phi(0)=0$
            is not the most natural choice since
            it would lead to the ill posedness of the
            Cauchy problem in the limiting case of small $\beta$.}.
The figure displays the phase evolution during 
{\em  3 sequential steps\/} of variation of $t$,
each of duration $T$.
The graph of the first period phase evolution is of red color.
The second period plot graph 
($t\in[T,2T]$) is brown.
To be placed over the
same $t$-axes segment as the first graph,
it
{\em is shifted to the left\/}
(`backwards in time')
just by $T$.
Similarly, the graph of the phase function
over the third period time step
($t\in[2T,3T]$, the green curve)
is shifted to the left
by $2T$ that places it over the same $t$-axes segment
$[0,T]$.
As a result, all the three subsequent 
portions of the phase evolution graph are displayed
over the common
segment of the $t$-axes
becoming more eligible for a visual collation
and the monitoring of the $T$-scale periodicity.

In passing,
it is worth emphasizing
that due to periodicity of $f$
the functions resulted from the above `translations'
by means of the $t$-shifts  aliquot $T$
{\em retain to
verify\/}  Eq.\ (\ref{eq1}).
Thus, applying such a trick,
the `contiguous' phase function
representing the phase evolution 
over
some time interval, 
perhaps semi-infinite, can be equally well represented by the
sequence
of phase functions (solutions of Eq.\ (\ref{eq1})) 
each defined on the finite
interval $[0,T]$.

Besides,
for the better clarity,
some additional artificial uniformly incrementing {\em vertical shifts\/}
are applied to separate
graph segments in Fig.\ \ref{f1}. 
The point is that we would like to recognize
the limiting curve to which the sequence
of the displayed ones
converges, and the details of such a conversion, provided it takes place.
Obviously,
it is reasonable
to add some uniformly incremented vertical space to the vertical positions
of separate graph segments
in order to visually distinguish the far
mutually close curves
which otherwise
would be seen in the plot overlaid in a messy way.
Specifically,
in the case of Fig.\ \ref{f1}, the
vertical shifts amounting to additional 0.04 units per each subsequent
graph segment 
are applied. This allows one to visually distinguish the second (brown) and the third
(green) phase graph segments showing simultaneously that the
corresponding functions are very close. In subsequent similar plots,
the benefit of the trick will be ever more clear.

It turns out that under the conditions assumed,
the phase function on
the third time step (the third period) apparently coincides
with the one on the second step
and thus already represents,
with accuracy characteristic for the plot resolution,
the desirable
limiting (asymptotic) phase time dependence.
The $T$-scale convergence proves here extremely fast.
It is also obvious 
that in view of the convergence of subsequent graph segments to the
common limiting curve
the values of $\phi$ on the
left and right graph boundaries of its domain
coincide, making evidence of the
periodicity of the limiting
solution of Eq. (\ref{eq1}) we deal with.

\begin{figure}
\includegraphics{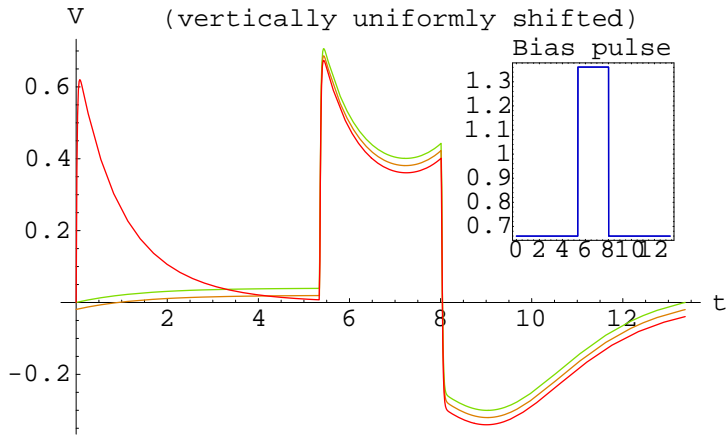}
\caption{\small
The voltage across junction (in arbitrary units) is displayed
for the phase functions
shown in Fig.\ \ref{f1}.
In order
to resolve the convergent
series of graphs,
the graphs except of the starting one
are uniformly shifted
in the `vertical' direction, the trick similar to one applied in Fig.\ \ref{f1}.
}\label{f2}
\end{figure}

 It is also of interest to consider the graphs of the
corresponding  {\em voltage\/} 
which coincides, up to the constant factor%
\footnote{ In order to evade introduction of dimensional numerics
which are irrelevant in the present discussion, we shall use throughout
instead the function $V$ as it is defined by Eq.\ (\ref{eq2}) the
`voltage function' $\dot\phi$. The plots of $\dot\phi$ will be
referred
to below as $V$ plots {\it in arbitrary units}.
},
with the derivatives of the
functions displayed in Fig.\ \ref{f1}.
These are shown in 
Fig.\ \ref{f2}, the vertical
and horizontal shifts (the part of {\it the plot segmenting\/} trick) similar to ones 
in Fig.\ \ref{f1} being applied. The voltage functions
also rapidly converge to a
periodic limiting one.

It has to be noted that the \phaselocking{}
does not necessarily claim
for the asymptotic
phase function $\phi(t)$ to be
periodic.
Indeed, 
keeping in mind the physical aspect of the problem, 
the
actually registered voltage across Josephson junction
is not momentary but is
efficiently
averaged over many bias periods $T$.
For the sake of simplicity,
let us consider the result of the voltage averaging
over a {\em single period} (the shortest time interval  making sense in the
situation under consideration), treating the  averaging 
as the plain
integrating 
followed by the division by the length of the integration interval.
Then in accordance with Eq.\ (\ref{eq2})
the averaging started at the moment $t$ has to yield
 \begin{equation}
\langle V\rangle_{\Delta t}= \Delta t^{-1}\int^{t+\Delta t}_{t} V(t')d t'\propto
\Delta t^{-1}[\phi(t+\Delta t)-\phi(t)]
                                          \label{volt}
\end{equation}
with $\Delta t=T$
(the proportionality relation means here the dropping out of a dimensional universal constant).
Thus the (asymptotic) periodicity of $\phi$ means 
$\langle V\rangle_{T}=0$ implying that
no average voltage across
the junction is observed%
\footnote{One of the forms of the Josephson effect:
some [averaged] bias current flows through the junction but the
[averaged] voltage across it 
is zero.
}.
However, instead of $f$ periodicity,
one may also assume that the
increment of $\phi$ across the time step of duration $T$
is independent
on the moment ${t}$ when the averaging starts:
$\phi({t}+T)-\phi({t})=\mbox{constant}$, asymptotically (for large
but otherwise arbitrary $t$).
This constant may not be arbitrary.
The condition above
is consistent with Eq.\ (\ref{eq1}) if, asymptotically,
 \begin{equation}
\phi({t}+T)-\phi({t})=2\pi k,
                                  \label{eq22}
\end{equation}
where $k$ is some integer, positive, negative, or zero.
This
number is named the {\em \phaselocking{}  order\/}.
Evidently, the
periodic phase function
is the particular case
corresponding to the zero order. Allowing arbitrary integer $k$,
the average voltage proves quantized 
assuming, in principle, any of discrete uniformly
                                distributed values%
 \footnote{More generally,
the average voltage quantization may also occur if $k$ is
a {\em rational\/} number,
provided the averaging is carried out over
sufficiently large number of periods.
}.

\begin{figure}
\includegraphics{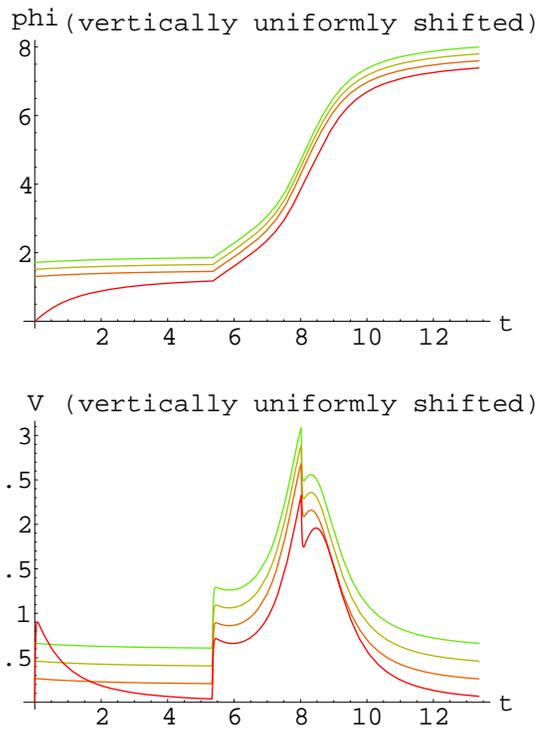}
\caption{\small
Example of the segmented phase function (top panel) and 
segmented
voltage function (lower panel, arbitrary units) for
the first order \phaselocking{} state.
The parameters of the model are the same as in
figures \ref{f1}, \ref{f2}, except for
$\iota_{\mbox{\scriptsize dc}}=1.1$.
The color of subsequent phase function segments uniformly varies
from red to green.
The uniformly incremented shifts
in the `vertical' direction are applied
including the first order \phaselocking{} responsible contribution
$-2\pi$ for the phase
($-2\pi$+0.2 units for the top panel and 0.2 units for the lower one
per subsequent graph segment,
respectively).
}
\label{f3}
\end{figure}

Numerical integration of Eq.\ (\ref{f1}) allows one
to easily confirm 
the feasibility
of non-zero-order
\phaselocking{} phase functions.
In particular,
Fig.\ \ref{f3} displays the example of {\em first order\/} \phaselocking.
 Here the only difference in the parameter
 setup with 
the zero order case
(figures \ref{f1}, \ref{f2})
is the value of the parameter
$\iota_{\mbox{\scriptsize dc}}$
now amounting to 1.1.
The limiting phase function is here {\em not\/} periodic
but its increments on the time steps of duration $T$ apparently 
tends to $2\pi$. On the other hand, subtracting the linearly growing
contribution which insures this secular phase incrementing, a
periodic function is leaved.

\begin{figure}
\includegraphics{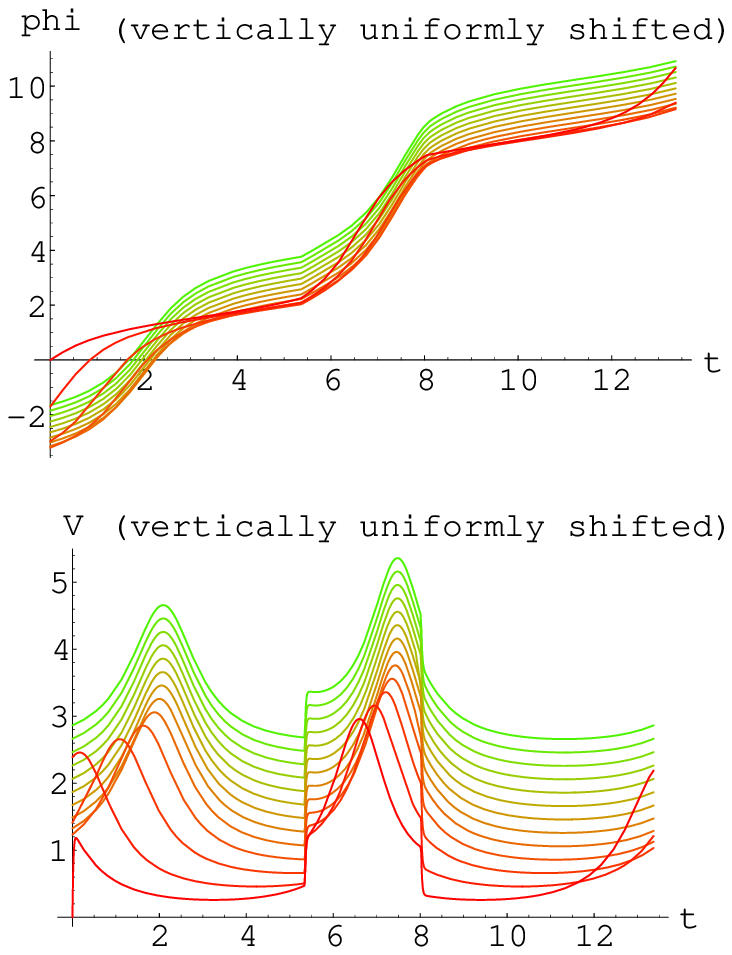}
\caption{\small
Example of the \phaselocking{} of the order two.
The parameters of the model are the same as in figures
\ref{f1}-\ref{f3}, except for
$\iota_{\mbox{\scriptsize dc}}=1.4$.
The color of subsequent phase function segments (top panel)
and voltage function segments (lower panel, arbitrary units)
uniformly varies
from red to green.
The uniformly incremented shifts
in the `vertical' direction
are similar to ones applied in Fig.\ \ref{f3} differing
in the shift contribution for the phase connected with the non-zero
order of the \phaselocking{} which amounts here to $-4\pi$ instead of $-2\pi$
applied therein.
}
\label{f4}
\end{figure}

Similarly, the further 
increasing of $\iota_{\mbox{\scriptsize dc}}$ yields
the second order
\phaselocking{} phase evolution.
It is displayed in Fig.\ \ref{f4}.
Here
$\iota_{\mbox{\scriptsize dc}}=1.4$,
the other parameters
being kept
unchanged.
Increasing $\iota_{\mbox{\scriptsize dc}}$,
the larger order \phaselocking{} phase functions can be produced as well.

\begin{figure}
\includegraphics{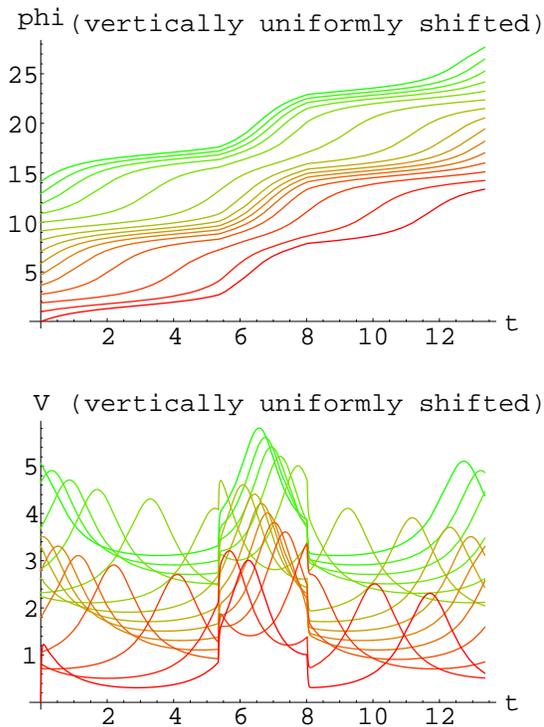}
\caption{\small
Example of asymptotically non-convergent ({{\chao}}) evolution.
The parameters of the model are the same as in figures
\ref{f1}-\ref{f4}, except of
$\iota_{\mbox{\scriptsize dc}}=1.45$.
The uniformly incremented shifts 
in the `vertical' direction ($-4\pi+0.2$ units for the phase 
and 0.2 units for the voltage functions
per subsequent graph segment, respectively)
are applied.
}                                   \label{f5}
\end{figure}

It is seen that in the
cases of higher order \phaselocking{},
the convergence
of the sequence of segmented phase functions
to the common limiting function
is slower.
Indeed, in the case of zero order
\phaselocking{} (Fig.\ \ref{f1},
$\iota_{\mbox{\scriptsize dc}}=0.8$)
the third
iteration apparently coincided
with the second one and yield,
in fact, the limiting curve
(the
\phaselocking{} order
can be estimated from the asymptotic relation
 \begin{equation}
 k\simeq
 (2\pi)^{-1}
 [\phi(t+T)-\phi(t)]
                                  \label{eq23}
 \end{equation}
taking place for sufficiently large $t$
(formally, in the limit $t\rightarrow\infty$), provided
the \phaselocking{} is developed).
In the case of the \phaselocking{} of the first
order
observed in the case
$\iota_{\mbox{\scriptsize dc}}=1.1$
the four $T$-lapses
are needed for the
reaching of
a comparable  closeness to the limit.
For the \phaselocking{} of the order two which takes place, in particular,
if
$\iota_{\mbox{\scriptsize dc}}=1.4$ the number
of necessary iterations increases up to 15.
And it suffices to set up
$\iota_{\mbox{\scriptsize dc}}=1.45$ to lose
the apparent convergence to any steady asymptotic
state
at all.
The corresponding apparently non-convergent segmented
phase and 
voltage function plots are shown in Fig.\ \ref{f5}.
These reveal no   tendency to converge
to any limiting curves
demonstrating
`irregular' (or even apparently
`chaotic', as far as one concerns the variations on the scale $T$) behavior.
The value of the expression (\ref{eq23}), irregularly
oscillating around
2, does not reveal apparent tending to any limit as well.

It is worth mentioning
that the convergence ---
or 
the divergence
---
of the sequence of phase functions
describing a single long-term phase
evolution
is often difficult
to reliably establish
by means of straightforward numerical integrating
of Eq.\ (\ref{eq1}).
Indeed, the slower
such a sequence converges, the longer
fragment of the phase evolution
has to be analyzed.
Accordingly, approaching the
hypothetical boundary of the
area in the space of the bias functions 
where the evolution of
the phase leads to a steady state
({\it \phaselocking{} area\/} in the parameter
space),
one inevitably encounters with
insufficiency of
the available computational resources.
The very boundary,
where the convergence reverses to the
divergence, is not detectable in this way.
Besides, 
in view of the obscurity of the corresponding time scale,
the `strong' phase divergence 
(more exactly, the absence of convergence to any asymptotic steady
state)
is
even harder to detect
through the apparent behavior of numerically generated phase function
than the too slow convergence.

The  formulation and substantiation
of the method allowing one to
efficiently  distinguish
the
\phaselocking{}
property 
in the important particular case of `overdamped'
version of equation (\ref{eq1})
is the main theme of the
present discussion.
The term `overdamped' refers here
the case 
 \begin{equation}
0\le\beta\ll1.
                                  \label{eq3}
\end{equation}
The parameter $\beta$
scales the (only)
term involving the second order derivative
of the phase function in  Eq.\ (\ref{eq1}).
It might be
supposed that under certain conditions
including imposing of limitation on the second derivative values
the relevant solutions of Eq.\
(\ref{eq1}) may be approximated
by the solutions of the equation
 \begin{equation}
  \dot \phi + \sin\phi=f
                                  \label{eq1X}
 \end{equation}
obtained from Eq.\
(\ref{eq1})
by a mere discarding of
the second order derivative term.

\begin{figure}
\includegraphics{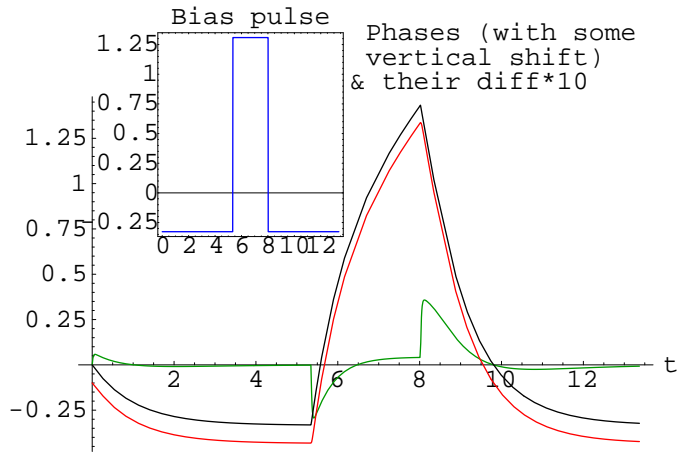}
\caption{\small
The related solutions of Eqs.\ (\ref{eq1})
(the red graph) and to 
(\ref{eq1X}) (the black  one)
for the same periodic rectangular pulse bias
(shown in inset)
displayed together with their
difference  magnified by the
factor of 10 (the green graph) .
The red plot is shifted downward by 0.1 units.
The bias is the same as above
except the parameter $\iota_{\mbox{\scriptsize dc}}$ which
is now assumed
to vanish and the pulse magnitude here equal to 3.5.
}
\label{f6}
\end{figure}

The numerical example illustrating 
the impact of the above `simplification' is shown in
  Fig.\ \ref{f6}, the
discussion
of the status
of the corresponding approximation for `overdamped' systems
(in the case of  sinusoidal bias $f$)
can be found in
\cite{X5}.

In Fig.\ \ref{f6}
the black  curve represents
the phase function computed in accordance
with  Eq.\ (\ref{eq1X}),
the red one is
the corresponding solution of the
general RSJ-equation 
(\ref{eq1}).
The `red' solution
is 
shifted downward by 0.1 units
in order to allow  easier visual distinguishing of the close graphs
on the common plot panel.
The green curve represents  the difference of the two
phase functions above 
which is {\em magnified by the factor of 10}.
The bias pulse is shown in the inset. Its form
is similar to the ones used above,
see Fig.\ \ref{f1}, except for
the bias `DC constituent',
$\iota_{\mbox{\scriptsize dc}}$,
which is here chosen vanishing (i.e.\  $\int^T_0 f({t})d\,{t}=0$),
and the integral  magnitude of a single pulse which here and in all
the numerical
examples below amounts to 3.5.

 It is seen that under the conditions assumed the
 solutions of
Eqs.\ (\ref{eq1X}) and (\ref{eq1}) are very
close, in total, both in shape and in
magnitude, the deviations emerging at several moments of time ultimately relaxing.
Specifically,
the solution difference 
reveals
the splashes originated in three points:
at ${t}=0$, 
and on the front and back `shocks' of the bias pulse.
At right near them, the difference soon
reaches a maximum and
then quickly die. In total, the distinction
of the corresponding solutions of Eqs.\ (\ref{eq1})
and (\ref{eq1X})
may be considered fairly
mild.

It is easy to see that the splash following the point ${t}=0$
arises due to the specific choice of the initial conditions
for solution of Eq.\ (\ref{eq1}) which read
$\phi(0)=0=\dot\phi(0)$. Then Eq.\ (\ref{eq1}) implies
$\ddot\phi(0)=\beta^{-1}f(0)$ which, for $f(0)\not=0$,
is a large quantity contrary to assumption required for approximating
 Eq.\ (\ref{eq1}) by (\ref{eq1X}).
In particular, it leads to the fast growth of the first derivative
immediately after the start of the phase evolution.
This effect is also observed in Fig.\ \ref{f2}.

Apart of the apprehensible effect of non-equivalent initial conditions,
the cause of the other discrepancy splashes is also quite understandable:
they arise due
to the discontinuities of the {\sc r.h.s.} function $f$.
Indeed, as a
pulse front/back jump occurs,
the discontinuity in $f$ converts  to a finite
jump in the derivative $\dot\phi$ in solution of Eq.\ (\ref{eq1X}) which, in turn,
emerges the Dirac $\delta$ function-like
irregularity in the second
derivative $\ddot\phi$ which again violate
the assumption
of limitation of $\ddot\phi$ magnitude
required for the approximating of solutions of (\ref{eq1})
by solutions of  Eq.\ (\ref{eq1X}).

Evidently, the cause of these peculiarities, 
leading to major distinctions of solutions of
 Eq.\ (\ref{eq1}) and Eq.\ (\ref{eq1X})
shown in Fig.\ \ref{f6},
originates
in either the admitting somewhat inadequate
initial conditions or in a
too rough representation
(by a discontinuous function) of the bias term.
Since
for a more physically realistic 
bias model 
no $\delta$-like
irregularities of the phase derivative can arise, the agreement of the
models based on
Eqs.\ (\ref{eq1}) and (\ref{eq1X}) would improve.

As compared to more deep Eq.\ (\ref{eq1}) whose study is still 
possible by
numerical methods alone,
Eq.\ (\ref{eq1X}) 
proves more 
transparent and allows fairly deep analytical treatment.
It seems also to be of
interest in its own rights
as the example of a non-linear problem of a deep physical relevance
associated
with a simple linear dynamical system%
\footnote{\label{ftr}
In the case of the bias function
$f(t)=B+A\cos\omega (t-t_0)$, where $A,B,\omega,t_0$ are some 
constant parameters, which is
most important from viewpoint of applications,
Eq.\ (\ref{eq1X}) proves converting to the following remarkable equation:
$$\left\{
(\zeta^2-1){\dex\over\dex \zeta}(\zeta^2-1){\dex\over\dex \zeta}
 - {2\over\omega}\left[
   (A+B) \zeta ^2
 + A-B
\right]
{\dex\over\dex \zeta}
+ \omega^{-2}\right\}\Upsilon=0
$$
}.
Eq.\ (\ref{eq1X}) 
possesses therefore a number of remarkable
properties which are the
subject of the discourse below.

\section*{The modeling of the phase dynamics
}

\subsection*{Master identity and its consequences}

\newtheorem{Mid}{Master identity}

The approach to the problem of
description of generic properties of solutions of Eq.\ (\ref{eq1X})
on the timescale exceeding the period $T$ including their asymptotic behavior
can be based on the following observation:
\begin{Mid}
For any piecewise smooth continuous functions
$\phi=\phi({t}), \phi_0=\phi_0({t})$,
let us define
the functions
 \begin{eqnarray}
P_0&\equiv&P_0({t})\equiv P_0[\phi_0]({t})=\int\cos\phi_0({t})\,d\,{t},
                                  \label{eq4} \\
Q_0&\equiv&Q_0({t})\equiv Q_0[\phi_0]({t})=\int e^{-P_0({t})}\,\sin\phi_0({t})\, d\,{t},
                                  \label{eq5}
 \end{eqnarray}
and introduce the notations:
 \begin{eqnarray}
\zeta&\equiv&\zeta({t})=e^{\imi\phi({t})},\enskip
                                  \label{eq6}
\zeta_0\equiv\zeta_0({t})=e^{\imi\phi_0({t})},
                                \\
C&\equiv&C[\phi,\phi_0]({t})=
\left[-Q_0+
\imi e^{-P_0}{\zeta_0+\zeta\over \zeta_0-\zeta}\right]^{-1}.
                                  \label{eq8}
 \end{eqnarray}
Then the following identity takes place
 \begin{equation}
{1\over 2 \imi}
(\zeta-\zeta_0)^2 e^{P_0} {d C^{-1}\over d\,{t}}\equiv
\zeta_0 D[\zeta] - \zeta D[\zeta_0]
                                  \label{eq9}
 \end{equation}
with the differential operator $D[\cdot]$ is defined as follows
 \begin{equation}
D[z]\equiv
 {d z\over d\,{t}} +{1\over2}(z^2 -1) -\imi f z
                                  \label{eq10}
 \end{equation}
for some function
$f\equiv f({t})$ and arbitrary piecewise smooth
$z\equiv z({t})$.
\end{Mid}

{\em Its proof\/} reduces to a straightforward computation
applying definitions of the functions involved.$\square$

\noindent
{\it Remarks:}
\begin{itemize}
\item
For the sake of definiteness, the indefinite integrals in
Eqs.\ (\ref{eq4},\ref{eq5}) can be understood
as $\int^{t}_0\dots d\,{t}$.
Another choice of lower integration boundary   would lead to
a {\em linear transformation\/} of $C[\dots]^{-1}$: an overall constant factor
is induced by the changing the lower boundary in the $P$-integral
while a  change of the lower boundary in $Q$-integral definition yields
a constant summand.
\item
Solving
Eq.\ (\ref{eq8})
with respect to $\zeta$, one gets the equation
 \begin{equation}
\zeta=\zeta_0
{1+C
(Q_0-\imi e^{-P_0})
\over
 1+C
(Q_0+\imi e^{-P_0})
}.
                                  \label{eq11}
 \end{equation}
Assuming $P_0,Q_0$ to be real,  for complex valued $\zeta_0,\zeta$
both of these variables can be
unimodular, $\vert\zeta_0\vert=1=\vert\zeta\vert$ (and $\phi_0,\phi$
real, see (\ref{eq6}
))
if and only if $C$ is real.
\end{itemize}

\newtheorem{coro}{Corollary}
\newtheorem{prop}{Proposition}

The connection of the relationships inferred
from the identity (\ref{eq9})
with the equation (\ref{eq1X})  follows from the 
identity
 \begin{equation}
 {d \zeta\over d\,{t}} +{1\over2}(\zeta^2 -1) -\imi f \zeta \equiv
\imi\zeta(\dot \phi + \sin\phi-f),
                                  \label{eq12}
 \end{equation}
provided $\zeta$ and $\phi$ are connected by Eq.\ (\ref{eq6}).
In conjunction with master identity,
it  immediately
 yields the following \cite{X7}
\begin{prop}\label{c1}
Let $\phi_0({t})$ verify Eq.\ (\ref{eq1X})
whereas $P_0({t}), Q_0({t})$ be defined by
Eqs.\ (\ref{eq4}), (\ref{eq5}).
Then any solution $\phi({t})$ of
Eq.\ (\ref{eq1X}) satisfies either the equation
 \begin{eqnarray}
e^{\imi\phi({t})}&=&e^{\imi\phi_0({t})}
{1+C(Q_0-\imi e^{-P_0})
\over
 1+C(Q_0+\imi e^{-P_0})
},
                                  \label{eq13}
 \end{eqnarray}
where $C$ is some real constant, or the equation
 \begin{eqnarray}
e^{\imi\phi({t})}&=&e^{\imi\phi_0({t})}
{
Q_0-\imi e^{-P_0}
\over
Q_0+\imi e^{-P_0}
},
                                  \label{eq13a}
 \end{eqnarray}
the latter
being in fact
the limiting form of the former as $C\rightarrow\infty$.
\end{prop}
{\it Remarks:}
\begin{itemize}
\item
When regarded as the relationship determining $\phi$,
Eq.\ (\ref{eq13}) or (\ref{eq13a}) specifies it not uniquely but
modulo $2\pi$. 
On the other hand, this is the only arbitrariness involved in such a
definition of $\phi(t)$.
\item
The initial, `ground' solution $\phi_0$ is produced
by the same
equation (\ref{eq13}) in the case
$C=0$ when (\ref{eq13})
converts to the identical transformation (modulo $2\pi$).
That is why just $C$ rather than $C^{-1}$ 
which has a formally simpler representation in terms of $\phi$'s, see (\ref{eq8}),
is employed in (\ref{eq13}).
\item
Given $\phi(0)$,
the constant $C$ can be easily found.
Indeed, it follows from Eq.\ (\ref{eq8}) that, placing the
left boundary of the integration interval at ${t}=0$,
one has  $P_0(0)=0=Q_0(0)$, that yields
 \begin{equation}
C=
\imi{
e^{\imi\phi(0)} -e^{\imi\phi_0(0)}
\over
e^{\imi\phi(0)}+e^{\imi\phi_0(0)}
}
=-\tan\half[\phi(0)-\phi_0(0)].
                                  \label{eq80}
 \end{equation}
Thus,
given any particular solution of Eq.\ (\ref{eq1X}) and
the integrals $P_0({t}), Q_0({t})$, defined from it by
Eqs.\ (\ref{eq4}), (\ref{eq5}) and obeying the conditions
$P_0(0)=0=Q_0(0)$, the solution of the Cauchy problem
for Eq.\ (\ref{eq1X})
is explicitly represented (modulo $2\pi$)
by Eqs.\ (\ref{eq13})  and (\ref{eq80})%
\footnote{Although the case $\phi(0)-\phi_0(0)=\pi\; \mathrm{mod}\, 2\pi$
is not covered, formally, by the above formulae,
a minor obvious modification  
corresponding to the transition to the limit $C\rightarrow\infty$
allows to treat it as well.
}.
\item
In view of (\ref{eq80}) one also gets
\begin{equation}
  \label{eq:urtcy}
  C[\phi_0,\phi]_{|t=0}=-C[\phi,\phi_0]_{|t=0}
\end{equation}
\end{itemize}

\begin{figure}
\scalebox{1.41}{
\includegraphics{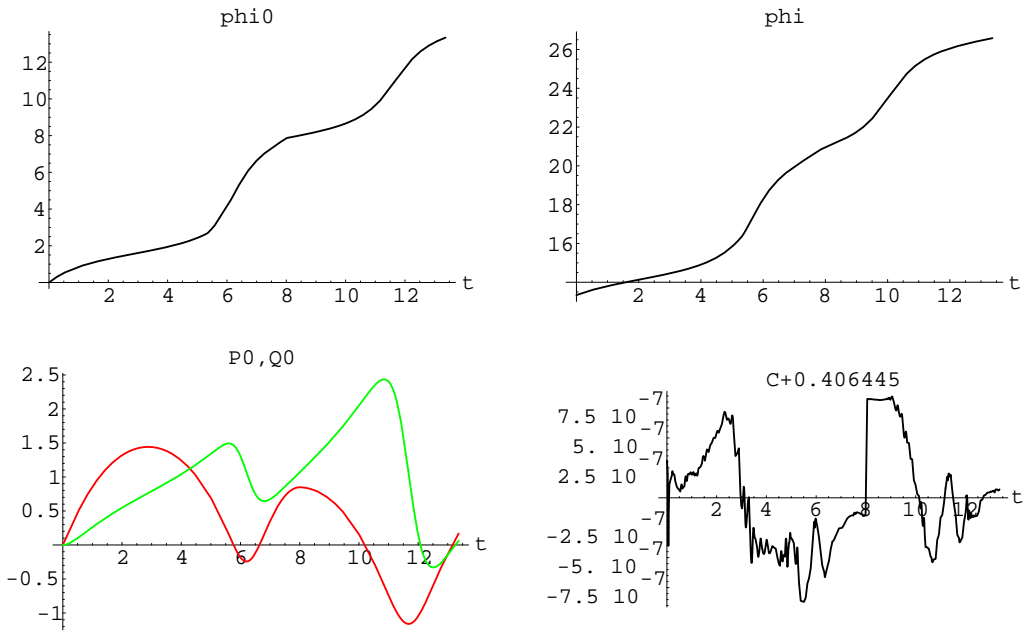}
}
\caption{\small
The top-left and top-right
panels display the phase functions $\phi_0({t})$, $\phi({t})$, respectively,
verifying
Eq.\ (\ref{eq1X}).
At the lower-left panel,
the integrals $P_0({t})$ (the red curve)
and $Q_0({t})$ (the green curve) are plotted,
see Eqs.\ (\ref{eq4}), (\ref{eq5}).
The deviation of
$C[\phi,\phi_0]({t})$ computed in accordance with
Eqs.\ (\ref{eq8}), (\ref{eq6}) 
from its average value
is displayed at
the lower-right panel.
}
\label{f8}
\end{figure}

Fig.\ \ref{f8} illustrates the above statements.
Here the top panels
contain the plots 
of the two different solutions of  Eq.\ (\ref{eq1X}).
The bias function $f$ is the same sequence of periodically repeated rectangular pulses
which was
used above, see the inset in Fig.\ \ref{f6},
the only difference is the constant constituent which 
now amounts to $\iota_{\mbox{\scriptsize dc}}=1.45$
(it is worth reminding also that now $\beta=0$).
The phase function of the first period phase evolution, starting at
the coordinate origin, is considered as the `ground' solution of
  Eq.\ (\ref{eq1X}) 
denoted above as
$\phi_0$.
It is plotted at the top-left panel.
In the lower-left  panel,
the integrals $P_0, Q_0$ calculated from $\phi_0$ are plotted.
The top-right panel displays the `continuation' of $\phi_0$ to the second
period $[T,2T]$ which, afterwards, is returned (`shifted' to the left by $T$)
to the $\phi_0$ domain. (No `vertical' shift is here applied.)
As it had been mentioned above,
due to the periodicity of $f(t)$, the
`shifted'
solution is again a
 solution of Eq.\ (\ref{eq1X}). We denote it $\phi(t)$. (This function
 is distinguished by the
specific initial condition  $\phi(0)=\phi_0(T)$ which however plays no
role in
the current context.)
Finally, the graph displayed in the lower-right panel
is the result of straightforward
computation of $C\equiv C[\phi,\phi_0]({t})$ in accordance with definition
(\ref{eq8}) and (\ref{eq6}). More exactly, the deviation of $C$ off its
averaged (on the interval $[0,T]$)
value which is $\approx$$-0.406445$
is shown.

One sees
that the 
functional $C=C[\phi,\phi_0](t)$ 
computed `from the first principles'
is, after all, a constant
up to a small deviation.
The latter 
irregularly oscillates around zero with the
amplitude which is at least 6 orders less than the phase magnitudes.
It  represents the
`numerical noise' reflecting mostly tolerable inaccuracy
of approximate numerical
solutions of differential
equation%
.

\noindent

\begin{figure}
\includegraphics{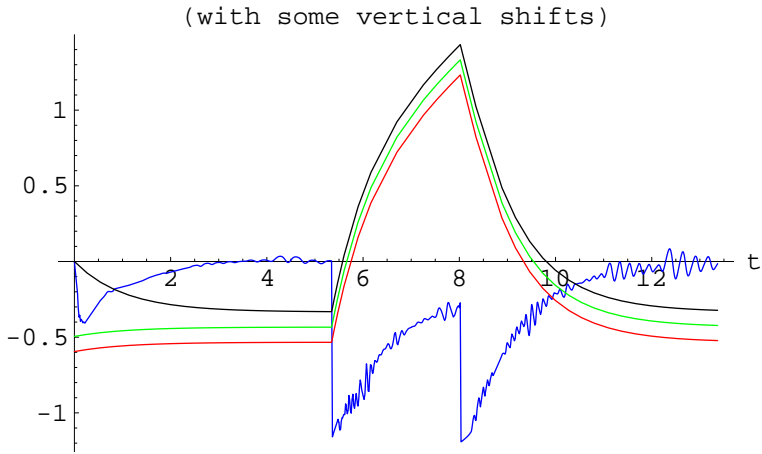}
\caption{\small
The black curve represents the `ground' phase function.
The red and green curves are the phase functions
(shifted downward by 0.1 and 0.2 units,
respectively) which are
computed in different ways:
the red curve is obtained by means of
Eq.\ (\ref{eq13}) for $C=0.2$
while the green one is the result of
numerical integration of Eq.\ (\ref{eq1X}) with  the same
initial conditions
as the `red' solution obeys.
Finally, the blue curve shows
the difference
of the ``red'' and ``green'' functions 
magnified by the factor of $0.5\times 10^6$.
}
\label{f9}
\end{figure}

The closely related computation displayed in Fig.\  \ref{f9}
illustrates the application of Eq.\ (\ref{eq13})
for the generating of new solutions of Eq.\ (\ref{eq1X})
from a known one.
Here the black curve shows the `ground' solution, $\phi_0$,
starting in the
coordinate origin; 
note that any other
phase function might be used instead.
Then we chose, say,
$C=0.2$ and compute $\phi({t})$ by means of Eq.\ (\ref{eq13}).
The result is represented by the red curve
which was
{\em shifted downward\/} by 0.1 units for the convenience 
of further visual collations.
Next, we carry out the straightforward
numerical integrating of Eq.\ (\ref{eq1X})
adopting
{\em the same initial conditions $\phi\vert_{{t}=0}=\phi(0)$} as the
just generated solution obeys.
The  green curve represents
the result of the integrating
which is also additionally
{\em shifted downward\/}, here by 0.2 units.
Finally, the blue graph is the difference of the
results of computation 
of the same phase function by these two methods
(viz the application of Eq.\ (\ref{eq13}) and the numerical ODE 
integrating) {\em magnified by the factor of  $0.5\times
  10^6$.} 
Again, one sees that Eq.\ (\ref{eq13}) 
ensures 
the stable accuracy 
about six true decimal digits
with the discrepancy
falling in the level of `numerical noise'.

It is important to note that, in the relationships above,
the roles of 
the functions $\phi_0$ and $\phi$
(any two solutions of Eq.\ (\ref{eq1X}))
should be symmetric  by general reasons.
However
Eq.\ (\ref{eq13}) does not reveal, apparently,
such a symmetry
since it involves the integrals
$P_0$, $Q_0$ determined by the solution $\phi_0$ but no similar
contribution
connected with $\phi$ is present.
Besides,
Eq.\ (\ref{eq13}) is easily solvable with respect to
$\phi$ but represents
to a nonlinear integral
equation  with respect to $\phi_0$.

Any inconsistency does not arises here however and 
the asymmetry mentioned above is actually
a fallacious one.
The point is that there is a 
specific algebraic relation
which constrains 
$P$- and $Q$-functionals associated with {\em any\/} two solutions
of Eq.\ (\ref{eq1X}).
It
enables one, in particular, to represent any $P,Q$  
as a simple
elementary function of another $P,Q$-pair.
Namely, the following remarkable relation takes place.
\begin{prop}
Let $\phi_0({t})$, $\phi({t})$, $P_0({t})$, $Q_0({t})$
be as in Proposition \ref{c1}.
Let also $P({t})$, $Q({t})$ be determined from $\phi({t})$
in the same way as $P_0({t})$, $Q_0({t})$ are
determined from $\phi_0({t})$. Then
 \begin{equation}
Q+\imi e^{-P}
=
{(Q_0+\imi e^{-P_0})-C
\over
 1+C(Q_0+\imi e^{-P_0})
}
                                  \label{eq14}
 \end{equation}
with {\em the same\/} real constant $C$.
\end{prop}
{\em Remarks:}
\begin{itemize}
\item Inverting (\ref{eq14}), one gets the alike formula
 \begin{equation}
Q_0+\imi e^{-P_0}
=
{(Q+\imi e^{-P})+C
\over
 1-C(Q+\imi e^{-P})
}
                                  \label{eq14a}
 \end{equation}
which differs from (\ref{eq14}), apart of interchanged roles of
$P,Q$ and $P_0,Q_0$, by the opposite sign of the $C$-constant alone.
\item From viewpoint of the induced transformation group structure, 
the transformation (\ref{eq14}) coincides with the relativistic
velocity addition rule.
\item
In view of the relationship above, it seems natural
to incorporate the real-valued functions
$P({t})$, $Q({t})$ into the complex-valued one defined as follows:
\begin{equation}\label{eq:yrtfers}
{\cal F}({t})=Q(t)+\imi e^{-P(t)}.
\end{equation}
Then 
the transformation described Eq.\ (\ref{eq14})
may be named, in a sense, meromorphic since ${\cal F}$ is expressed
via ${\cal F}_0$ as  a (meromorphic) function of ${\cal F}_0$.
\item
The definition of $P,Q$-integrals  by
(\ref{eq4}), (\ref{eq5})-like equations is equivalent to
the equation
\begin{equation}
{\dex\over \dex t}{\cal F}(t)=-\imi e^{\imi\phi(t)}\Im{{\cal F}(t)}.
                                        \label{eq85}
\end{equation}
Indeed, rewriting it as
${\dex(\Re{\cal F})/ \dex t}+\imi{\dex(\Im{\cal F})/ \dex t}=
(\sin\phi-\imi\cos\phi)(\Im{\cal F})$
and separating the pure imaginary part, one gets
${\dex(\Im{\cal F})/ \dex t}=-\cos\phi\cdot\Im{\cal F}$,
which, for nonzero  $\Im{\cal F}$, yields
$\Im{\cal F}=\exp(-\int\cos\phi \dex\phi)$,
i.e. in accordance with definition (\ref{eq:yrtfers}), Eq.\ (\ref{eq4}) in fact.
Next, separating the real part, one gets
${\dex(\Re{\cal F})/ \dex t}=\sin\phi\cdot\Im{\cal F}$.
Integrating it and taking into account the representation of
$\Im{\cal F}$ just obtained, one gets Eq.\
 (\ref{eq5}).$\square$
\item
Working in terms of $\cal F$ instead of $P,Q$,
the choice of $t=0$ as the lower integral limit in  (\ref{eq4}),
(\ref{eq5}),
is equivalent to the initial condition
\begin{equation}
{\cal F}(0)=\imi
                                        \label{eq86}
\end{equation}
for the function $ {\cal F} $ verifying  Eq.\ (\ref{eq85}).
\item
Inverting Eq.\ (\ref{eq14}),
${\cal F}_0$ can be represented as a function of ${\cal F}$ (see (\ref{eq14a})).
Then  Eq.\ (\ref{eq1X}) can be 
solved with respect to $\phi_0(t)$ which is represented
as explicit function of $\phi(t)$, $P({t})$, $Q({t})$ (and $C$) as follows
 \begin{eqnarray}
e^{\imi\phi_0({t})}&=&e^{\imi\phi({t})}
{1-C(Q-\imi e^{-P})
\over
 1-C(Q+\imi e^{-P})
},
                                  \label{eq13b}
 \end{eqnarray}
The dual Eqs.\ (\ref{eq13}),(\ref{eq13a})
manifest
the symmetric role of
the two solutions 
$\phi,\phi_0$ noted above.
\end{itemize}
{\it Proposition proof}.
The relationship to be proven belongs to the category of ones 
for which, as the true formula is recorded, 
the proof reduces to
a straightforward computation.
Indeed,
let us rewrite Eq.\ (\ref{eq13}) as follows
\begin{equation}\label{eq:ogytgfdr}
e^{\imi\phi(t)}=e^{\imi\phi_0(t)}
{1+C\overline{{\cal F}_0}
\over
 1+C{{\cal F}_0}
}
\end{equation}
and calculate the $t$-derivative of the difference
$$
\delta\equiv {\cal F}-{{\cal F}_0-C\over1+C{\cal F}_0}=
{\cal F}-{1\over C}+{1+C^2\over C(1+C{\cal F}_0)}.
$$
Then the straightforward computation yields
\begin{eqnarray}
{\dex\delta\over\dex t}&=&
{\dex{\cal F}\over\dex t}-
{1+C^2\over (1+C{\cal F}_0)^2}{\dex{\cal F}_0\over\dex t}
                        \nonumber\\
&=&
\half e^{\imi\phi}(\overline{\cal F}-{\cal F})-
\half {1+C^2\over (1+C{\cal F}_0)^2}e^{\imi\phi_0}
(\overline{\cal F}_0-{\cal F}_0)
                        \nonumber\\
&=&
\half e^{\imi\phi_0}\left(
{1+C\overline{{\cal F}_0}\over 1+C{{\cal F}_0}}
(\overline{\cal F}-{\cal F})-
{1+C^2\over (1+C{\cal F}_0)^2}(\overline{\cal F}_0-{\cal F}_0)
\right).
                        \nonumber
\end{eqnarray}
The following identity takes place
\begin{eqnarray*}
\overline{\cal F}-{\cal F}
&=&\overline{\delta}-{\delta}
-
{1+C^2\over C}\left(
\overline{1\over1+C{\cal F}_0}-{1\over1+C{\cal F}_0}
\right)
                        \nonumber\\
&=&\overline{\delta}-{\delta}
+
{1+C^2
\over
({1+C{\cal F}_0})
{(1+C\overline{\cal F}_0)}
}
(\overline{{\cal F}_0}-{\cal F}_0).
\end{eqnarray*}
It implies, together with the equation just derived, the following 
series of equalities:
\begin{eqnarray*}
{\dex\delta\over\dex t}&=&
\half e^{\imi\phi_0}\left(
{1+C\overline{{\cal F}_0}\over 1+C{{\cal F}_0}}
\left(
\overline{\delta}-{\delta}
+
{1+C^2
\over
({1+C{\cal F}_0})
{(1+C\overline{\cal F}_0)}
}
(\overline{{\cal F}_0}-{\cal F}_0)
\right)
\right.
                        \nonumber\\
&&\left.\hspace{5ex}
-{1+C^2\over (1+C{\cal F}_0)^2}(\overline{\cal F}_0-{\cal F}_0)
\right)
                        \nonumber\\
&=&
\half e^{\imi\phi_0}\left(
{1+C\overline{{\cal F}_0}\over 1+C{{\cal F}_0}}
(
\overline{\delta}-{\delta})
+
{1+C\overline{{\cal F}_0}\over 1+C{{\cal F}_0}}
{1+C^2
\over
({1+C{\cal F}_0})
{(1+C\overline{\cal F}_0)}
}
(\overline{{\cal F}_0}-{\cal F}_0)
\right.
                        \nonumber\\
&&\left.\hspace{5ex}
-{1+C^2\over (1+C{\cal F}_0)^2}(\overline{\cal F}_0-{\cal F}_0)
\right)
                        \nonumber\\
&=&\half e^{\imi\phi_0}
{1+C\overline{{\cal F}_0}\over 1+C{{\cal F}_0}}
(\overline{\delta}-{\delta}).
\end{eqnarray*}
The last equation is equivalent to the system of two linear homogeneous
first order ODEs for the two (real valued) functions $\Re\delta,\Im\delta$.
Moreover, in accordance with standard `normalization'
(\ref{eq86}) and $\delta$
definition, one has
\begin{equation}
\delta(0)=0,
                \nonumber
\end{equation}
the null initial condition.
Thus $\delta(t)\equiv 0$, i.e.\ one gets the equation
\begin{equation}
{\cal F}={{\cal F}_0-C\over1+C{\cal F}_0}
\end{equation}
which is nothing but Eq. (\ref{eq14}). $\square$

Thus, in accordance with the remarks above, we may consider
the expression (\ref{eq8}) as the functional over the
set of pairs of solutions of Eq.\ (\ref{eq1X})
(the direct square of the solution set)
taking
it onto the real axis. To be more exact,
since the limiting case
$C\rightarrow\infty$
is quite legitimate, one has to
replenish
the real axis 
by the infinitely remote point.
The result is the
circumference but it appears here as the isomorphic
image of the real projective line $RP^1$ which is just the
natural `index space' to be used for the `enumerating' of
solutions of Eq.\ (\ref{eq1X}).
Specifically,
having fixed some $\phi_0$,
the inverse map from $S^1\simeq RP^1$ onto the first factor in the
direct product
yields the (1-1 non-canonical) 
parameterization of the space of
solutions by the circumference points.

It is worth noting that
the $C$-map (\ref{eq8}), (\ref{eq6})
is {\em antisymmetric\/}
with respect to the interchange of the functional arguments $\phi,\phi_0$, i.e.\
 \begin{equation}
C[\phi,\phi_0]=-C[\phi_0,\phi].
                                  \label{eq15}
 \end{equation}
The property
(\ref{eq15})
can be established as follows.
Eq 
(\ref{eq9}) implies
 \begin{equation}
e^{P_0} {d C[\phi,\phi_0]^{-1}\over d\,{t}}=
 2 \imi(\zeta-\zeta_0)^{-2}(\zeta_0 D[\zeta] - \zeta D[\zeta_0])=
-e^{P} {d C[\phi_0,\phi]^{-1}\over d\,{t}}
\nonumber
 \end{equation}
and therefore
$C[\phi_0,\phi]$ is automatically a constant, provided 
$C[\phi,\phi_0]$ is.
Further, in accordance with
(\ref{eq:urtcy})
the equality
$-C[\phi_0,\phi]=C[\phi,\phi_0]$ takes place for $t=0$.
Hence these constants coincide up to the opposite signs and
(\ref{eq15}) holds everywhere. 
\begin{prop}
If $\phi,\phi_0$ are solutions of Eq (\ref{eq1X})
then Eq.\ (\ref{eq15}) is satisfied. $\square$
\end{prop}

\subsection*{Phase-locking and its necessary condition}

Now let us consider the implications of the above relationships
in application to
the property of the \phaselocking{}.
The latter is connected
with the  specific asymptotic  behavior of the phase 
functions
$\phi({t})$ verifying, in our case,  Eq.\ (\ref{eq1X}).
Namely,
in the case of \phaselocking{}
Eq.\ (\ref{eq22}) has to be satisfied,  asymptotically.

To describe efficiently this property,
let us define the sequence of functions $\phi_j(t)$ 
defined on the segment $[0,T]$
as follows%
\footnote{
There is some abuse in these notations prone of a  mix of
the meanings of the symbol $\phi_0$. 
Above, it was understood as one of the phase functions from the pair $\phi,\phi_0$
(see e.g.\ (\ref{eq6})) or as the `ground' solution obeying 
the 
condition $\phi_0(0)=0$
as in Fig.\ \ref{f8}, such a usage leading to no interpretation problems. 
The new interpretation of the same symbol
is the particular element of the sequence of functions 
defined by Eq.\ (\ref{eq16}). It does not match the above. 
Usually, it is clear what is meant. However,
if below the both meaning loads of 
the symbol $\phi_0$ (as well as $P_0$, $Q_0$) meet in a common context, 
a slightly modified 
notation, $\phi_{(0)}$, will
be employed for its first interpretation.
} 
\begin{equation}
\phi_j({t})=\phi({t}+j T) - 2\pi[[\phi(j T)/2\pi]],
\;j=0,1,2,3\dots,\; {t}\in{}[0,T]
                                  \label{eq16}
 \end{equation}
where the double brackets $[[\dots]]$ stands for
the integer part of the real number enclosed.
Thus the plot of $\phi_j({t})$
displays 
how the `genuine' phase function
$\phi({t})$ looks like `on the $j$'th segment' of the length
(duration) $T$
with respect to the closest level
aliquot to $2\pi$.
To that end,
its graph is
shifted downward or upward
by such a number of  `full phase revolutions'
$2\pi$ 
which returns 
its left-boundary point $\phi_j(0)$ to the segment
$[0,2\pi)$.

We have already used such a  trick 
in the arrangement of plots of phase functions
calling it occasionally `the segmenting'. 
Here the explicit
transformation of the phase function on the corresponding
t-segments of duration $T$ 
yielding a specific sequence of the phase functions defined on the segment $[0,T]$
is introduced.


All the functions $\phi_j$ satisfy Eq.\ (\ref{eq1X}).
The
specific (and
characteristic)
property of the
sequence of its solutions $\phi_j$'s
associated with a {\em single\/} solution $\phi$ defined for all $t$
is obviously the following:
 \begin{equation}
  \phi_{j+1}(0)=\phi_j(T)\mbox{\ mod $2\pi$}, \phi_{j}(0)\in[0,2\pi), j=0,1,2\dots
                                \label{eq16a}
 \end{equation}
Given the sequence $ \{ \phi_j \}$ of solutions of Eq.\ (\ref{eq1X})
on the segment $[0,T]$
fulfilling the condition (\ref{eq16a}), the valid phase function
$\phi(t), t\in[0,\infty),$ can
be reconstructed in the obvious way.

It is convenient to represent the \phaselocking{} property
in terms of properties of the functional sequence
$\phi_j$. 
It can be stated the following:
\begin{quote}
In the case of \phaselocking{} the sequence
$\phi_j$ uniformly converges on the segment $[0,T]$ to some limiting function
 \begin{equation}
\phi_\infty({t})= \lim_{j\rightarrow\infty} \phi_j({t})
                                  \label{eq17}
 \end{equation}
which satisfies Eq.\ (\ref{eq1X})
and the condition
 \begin{equation}
\phi_\infty(T)=\phi_\infty(0)+2\pi k
\Leftrightarrow
e^{\imi\phi_\infty(T)}=e^{\imi\phi_\infty(0)}
                                  \label{eq18}
 \end{equation}
for some integer k.
\end{quote}

It is natural to select and fix the `ground' specimen
of phase functions denoting it $\phi_0({t})$ , ${t}\in{}[0,T],$ characterizing it
by means of the null initial value $\phi_0(0)=0$
(the symbol
$\phi_{(0)}({t})$ and its derivates
will be also used in cases prone of confusion 
with the notation introduced in (\ref{eq16})). 
Having posed the problem in
such a way,
$\phi_0({t})$
is completely
determined by the bias function $f({t})$.
We also assume the value ${t}=0$ to be the lower integration boundary
for the integrals in the
$P_0, Q_0$ definitions  (\ref{eq4}),(\ref{eq5}).
Then
Eq.\ (\ref{eq13}) implies that
there exists a  constant $C_\infty$ such that
 \begin{equation}
e^{\imi\phi_\infty({t})}=e^{\imi\phi_0({t})}
{1+C_\infty(Q_0({t})-\imi e^{-P_0({t})})
\over
 1+C_\infty(Q_0({t})+\imi e^{-P_0({t})})
}.
                        \label{eq181}
 \end{equation}
(We assume $C_\infty$ to be finite but the case of `infinite $C_\infty$'
is also tractable, with minor modifications, in the same way.)
Taking the above representation of $e^{\imi\phi_\infty({t})} $ into
account,
it is easy to see that
Eq.\ (\ref{eq18}) is satisfied if and only if
the equation
 \begin{eqnarray}
&&
    C_\infty^2
\times
(Q_0(T)
 \cos\half \phi_0(T)
+ e^{-P_0(T)}
 \sin\half \phi_0(T))
                        \nonumber\\
&&
\llap{+}
    C_\infty
\times
(
   Q_0(T)  \sin\half\phi_0(T)
  +(1-e^{-P_0(T)})
\cos\half\phi_0(T)
)
                        \nonumber\\
&&
\hphantom{C_\infty}
+
\sin\half\phi_0(T)=0
                                  \label{eq19}
 \end{eqnarray}
admits a {\em real\/} root $C_\infty$. In turn,
the latter condition is equivalent
to the inequality
 \begin{eqnarray}
\Delta[f]&\equiv&
-Q_0(T)
(e^{-P_0(T)}+1)
\sin\phi_0(T)
                        \nonumber\\
&&
-\half
\left(-Q_0(T)^2+\left(e^{-P_0(T)}+1\right)^2\right)
(1-\cos\phi_0(T))
                        \nonumber\\
&&
+(1-e^{-P_0(T)})^2
                        \nonumber\\
&\equiv&
4 e^{-P_0(T)}\times
                        \label{eq20a}\\&&
\left(
 \left[
-\half
 e^{\half P_0(T)} Q_0(T)\sin\half\phi_0(T)
 +\cosh\half{}P_0(T)
 \cos\half\phi_0(T)
\right]^2
 -1
\right)
                                \nonumber\\
 &\ge&0
                                \label{eq20}
 \end{eqnarray}
 \begin{eqnarray}
   \label{eq:yrtred}
   \Leftrightarrow
&\left|
 \cosh\half P_0(T)
 \cos\half\phi_0(T)
-\half
 e^{\half P_0(T)} Q_0(T)\sin\half\phi_0(T)
\right|\ge 1.
&
 \end{eqnarray}
The above is therefore the {\em necessary\/} condition of
\phaselocking. 
It reflects in fact the property of the very
function $f$.

As we shall show,
the similar but {\em strict} inequality
is the {\em sufficient\/} condition of the \phaselocking{} for
the phase function described by Eq.\ (\ref{eq1X}).
(In the case of the equality, the convergence to some asymptotic limit
is observed as well but it is slower and reveals some other specialities.)
The vantage 
of such a form of the \phaselocking{} `monitoring' is that
this is  the {\em asymptotic\/} property
manifesting itself for sufficiently large $t$.
Generally speaking, one cannot forecast in
advance how long phase evolution has to be tracked
for the detection of the corresponding $T$-scale reproducibility of the phase
function form 
which would
make evidence of the \phaselocking{}.
On the other hand,
making use of the condition (\ref{eq:yrtred}),
the appearance of \phaselocking{}
is deduced directly from the properties
of a {\em single\/} solution of Eq.\ (\ref{eq1X})
computed on the {\em finite\/} interval $[0,T]$.

\subsection*{More on the role of the discriminant $\Delta[f]$ }

The function $\Delta\equiv\Delta[f]$ plays an important role 
in 
the problem 
of description of asymptotic properties of solutions of Eq.\ (\ref{eq1X}).
We may interpret it as the functional
on the set of bias functions $f$
since the functions $\phi_0, P_0, Q_0$ from which it is
built upon are uniquely determined by $f$: $\phi_0$ is
the solution of Eq.\ (\ref{eq1X}) with initial condition
$\phi_0(0)=0$, $P_0, Q_0$ are calculated from $\phi_0$
in accordance with
Eqs.\ (\ref{eq4}), (\ref{eq5}) and the initial conditions
$P_0(0)=0, Q_0(0)=0.$
One may also choose $f$ to belong to some more restricted class of
functions, for example, a family
parameterized by a finite set of parameters.
Then,
one may
also regard $\Delta$ as the function of these parameters and study
its properties.

\begin{figure}
\includegraphics{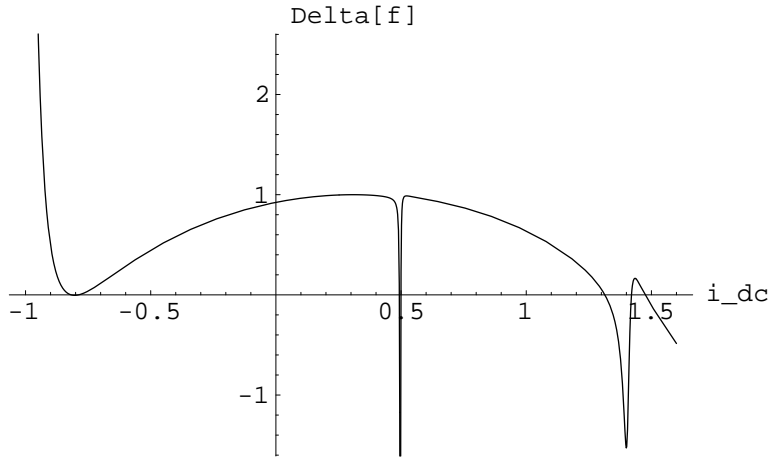}
\caption{\small
The dependence of the discriminant $\Delta$ on the constant
constitution $\iota_{\mbox{\scriptsize dc}}$ 
of the bias with the other parameters held fixed.
}
\label{f10}
\end{figure}

Adopting the last interpretation,
Fig.\ \ref{f10} shows 
$\Delta$
%
as the function of a single variable%
\footnote{We omit here and below the `functional argument' $[f]$ of $\Delta$,
provided this cannot lead to a misunderstanding.}, 
the constant bias constituent
$\iota_{\mbox{\scriptsize dc}}$,
assuming the very bias function to be the periodic sequence of
rectangular pulses
shown in the inset in Fig.\ \ref{f6}.

\begin{figure}
\scalebox{1.41}{%
\includegraphics{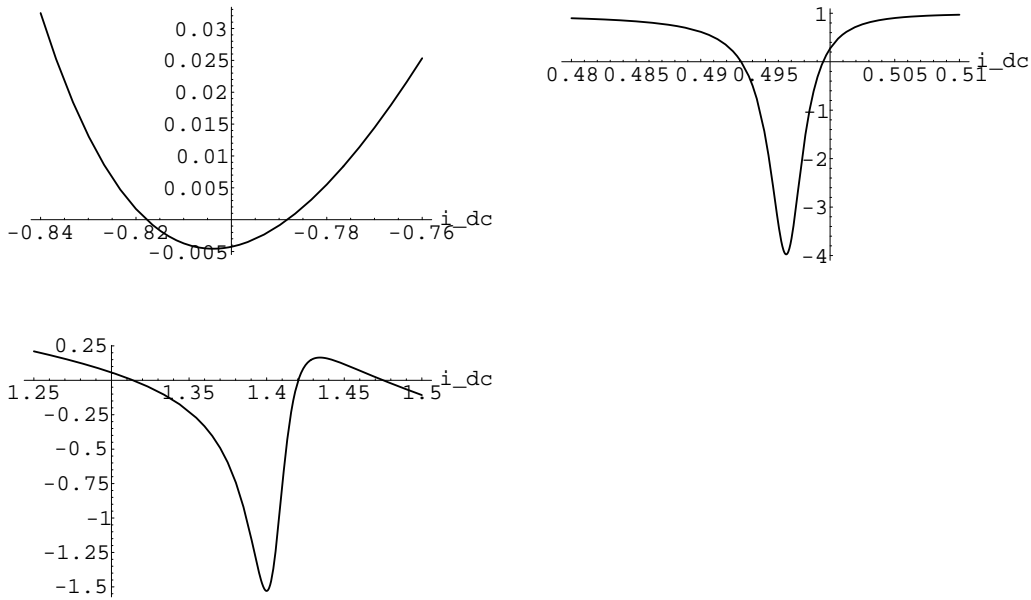}
}
\caption{\small
Separate
representations
of vicinities of the minima observed
on the
graph shown in Fig.\ \ref{f10}
are given with higher plot resolutions.
}
\label{f11}
\end{figure}

One recognizes the three minima on
the fragment of the $\Delta$ plot seated in Fig.\ \ref{f10},
the left one situating
very close to the horizontal coordinate axis and the middle one being fairly steep.
More  narrowly, the minima above are shown
in Fig.\ \ref{f11}, where the plots of their vicinities
are displayed with higher resolutions.
One sees that in all the three cases the minima lay well below the horizontal
coordinate 
axes. Hence each of them is encompassed by some $\idc$
segment
where $\Delta$ assumes negative values (and this is the universal
property: there are no positive minima of $\Delta$).
These are
separated by segments where $ \Delta$
is positive.

As we shall show,
each segment of positive $\Delta$ 
determines the area of the values of the parameter
$\iota_{\mbox{\scriptsize dc}}$ giving rise to the \phaselocking{}
phase evolution
{\em of a common order}.
Such segments are directly associated
with so called Shapiro steps --- strictly constant voltage segments
observed on the Josephson
junction I-V curve  \cite{X6,X4}.
 In particular, the domain of $\iota_{\mbox{\scriptsize dc}}$ values
shown in Fig.\ \ref{f10},
where $\Delta>0$,
contain the steps of the orders, from left to right, $-1$ (extending
to the left beyond the plot boundary), 0, 1, and 2.

The values of $\iota_{\mbox{\scriptsize dc}}$ falling into the segments
of negative $\Delta$ correspond to the apparently {{\chao}}
behavior of the
phase which is
similar to one displayed in Fig.\ \ref{f5}.
We shall see however that these phase evolutions do not 
correspond to  an actual chaos (pseudo-chaos) \cite{X5}.
 Rather, these are
the manifestations of a `beating' produced by two inconsonant
frequencies.
In particular, such phase evolutions
imply quite definite average voltages 
which are obtained by the averaging of (\ref{volt})
over large time intervals.
This point  will be considered in more details later on.

\subsection*{Unstable and stable \phaselocking{} candidates}

The
algorithm allowing one to reconstruct
the phase functions
at any moment of time $t$ 
divides into several steps.
At first,
one has to  solve the quadratic equation (\ref{eq19}) with respect to $C$,
assuming its roots 
to be real (that takes place iff $\Delta\ge0$).
Then, if $\Delta>0$,
Eq.\ (\ref{eq13}) yields the {\it two\/}
phase functions 
which are 
the candidates to the role of possessor
of the `refined' property of
\phaselocking{} 
(the limiting function asymptotically approximating generic solutions).

A somewhat delicate point is the revealing which of the two
solutions 
is the one we search for.
The answer is connected with the stability property 
(the \phaselocking{} must be  stable)
and can be provided, 
in principle, with the help of the standard perturbational
analysis (the local stability problem, cf.\ \cite{X5}) which however
yield not the actual problem 
solution but rather a recipe of its computation.
Fortunately,  there is an attractive opportunity
to directly manifest not only the effect of {\it infinitesimal\/} perturbations
but, at one blow,  to  establish the 
`global' stability property allowing perturbations of arbitrary
magnitudes.
Here we seize upon it.

\begin{figure}
\includegraphics{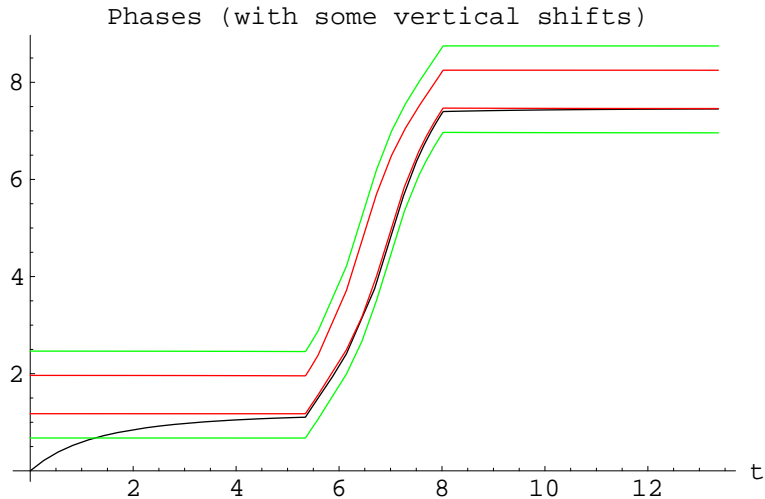}
\caption{\small
The black curve shows the `ground' solution $\phi_0$
(starting from the coordinate origin). The two red curves are the
the candidates to 
the role of the asymptotic limiting phase function 
which are constructed
by means of  Eqs.\ (\ref{eq19}), (\ref{eq13}).
The green curves are the phase functions constructed
by means of a straightforward numeric integration of (\ref{eq1X})
obeying  the same initial conditions as
the `red' solutions;
for better clarity,
they are shifted, afterwards, in `vertical' direction
by 0.5 units
upward (the top curve) and downward (the lower curve),
the red graphs being undergone no shifts.
}
\label{f12}
\end{figure}

\begin{figure}
\includegraphics{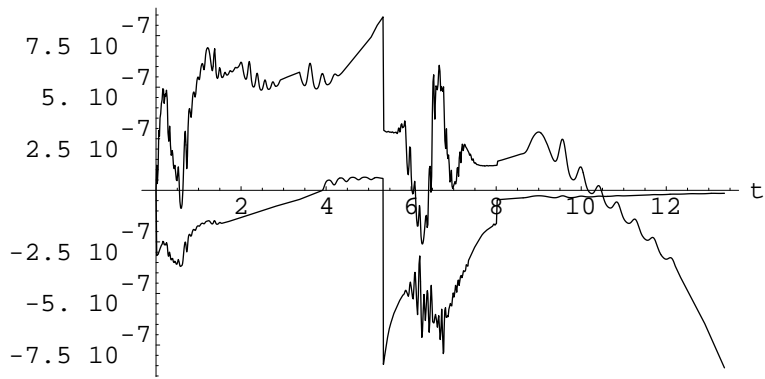}
\caption{\small
The plots display
the discrepancies between the \phaselocking{} candidates
displayed in Fig.\ \ref{f12} and obtained
by means of Eqs.\ (\ref{eq19}), (\ref{eq13}) 
against the ones
obtained
by means of straightforward
numerical integration of Eq.\ (\ref{eq1X})
with the same initial conditions as the former functions obey.
}
\label{f13}
\end{figure}

This point is illustrated by Fig.\ \ref{f12}.
Here the
black curve shows the `ground' solution $\phi_0$
(starting from the coordinate origin) 
for the rectangular pulse bias of the profile shown
in the inset in Fig.\ \ref{f6},
and $\iota_{\mbox{\scriptsize dc}}=1.25$ (which yields a positive
$\Delta$ value, see Fig.\ (\ref{f10})).

The red curves are the graphs of 
solutions  (`\phaselocking{} candidates') one of which is expected to
realize the `refined', precisely steady \phaselocking{}  evolution
obeying the
property (\ref{eq22}) and representing, in a sense,
the asymptotic limit of a generic phase function.
The \phaselocking{} candidates are constructed
by means of  Eqs.\ (\ref{eq13})
with the two values of the constant $C$ obtained from
Eq.\  (\ref{eq13}).

The green curves represent the phase functions constructed
by means of a  straightforward numeric integrations of Eq.\ (\ref{eq1X})
obeying  {\it the same initial conditions\/} as
the `red' solutions does. The plot graphs are afterwards
shifted `in vertical direction'  by 0.5 units
upward (the green top curve) and downward (the lower one)
making easier the visual collation
of the  profiles of the functions involved.

The apparent qualitative coincidence 
of the results of the two ways of computation of `\phaselocking{} candidates'
is numerically confirmed
in Fig.\ \ref{f13} where
their relative discrepancies are plotted.
One sees that the two computations
agree at the level one part in $10^{6}$ which is 
close to the accuracy of the numerical integrating of the ODE involved.

\begin{figure}
\includegraphics{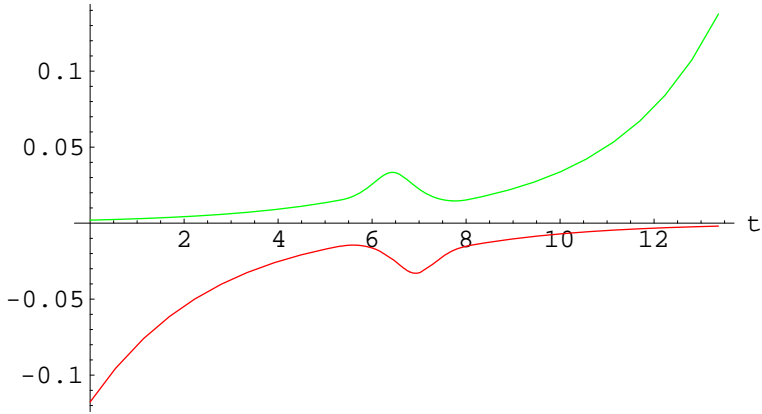}
\caption{\small
The evolution of perturbations of the two candidates 
the
`refined' \phaselocking{}
states represented in Fig.\ \ref{f13} by red graphs are displayed,
the perturbation
function corresponding to the stable evolution (the red graph)
being
magnified by the factor of $10^2$.
}
\label{f14}
\end{figure}

It is natural to suppose that only one of the two solutions
shown as the red (or, equivalently, green) curves in Fig.\ (\ref{f12})
is of interest
as a model of physical relevance
because another one
is necessarily {\em unstable\/} 
(it plays the own specific role 
as the separator of two `basins' of the \phaselocking{} 
in the space of all phase
functions, though).
Indeed, the distinction of their stability properties 
is clearly seen in Fig.\ \ref{f14}.

Here
the evolution of perturbations of the two solutions
shown in Fig.\ \ref{f13} (the red curves) is displayed,
The perturbed phase functions are defined as the
solutions $\phi^{(\varepsilon)}({t})$ to Eq.\ (\ref{eq1X})
with initial
conditions distinct from the ones
the \phaselocking{} candidates obey
by amount of 0.1\%:
$\phi^{(\varepsilon)}(0)=\phi_{}(0)\times(1\pm0.001)$.
For definiteness, we chose
$\phi^{(\varepsilon)}(0)=\phi_{}(0)\times(1+0.001)$
for the upper
curve in Fig.\ \ref{f13}  and
$\phi^{(\varepsilon)}(0)=\phi_{}(0)\times(1-0.001)$
for the lower one.
Accordingly, the upper (green) graph in Fig.\ \ref{f14}
shows $\phi^{(\varepsilon)}({t})-\phi_{}({t})$
for the `upper' \phaselocking{} candidate;
it makes evidence of a strong instability since
the deviation is permanently
(and, in fact, exponentially) growing.
On the contrary, the lower (red) curve in Fig.\ \ref{f14}
demonstrate the exponential relaxation of the perturbation;
moreover, to display it more clearly,
the perturbation value displayed is
magnified by the
factor of $10^2$, i.e.\ the function
$100(\phi^{(\varepsilon)}({t})-\phi_{}({t}))$
is actually plotted.

Thus we see that only the lower red curve in Fig.\ \ref{f13}
corresponds to
a stable phase evolution.
The evolution described by another, upper, (red) curve in Fig.\ \ref{f13}
related to
another solution of Eq.\ (\ref{eq19}), is unstable.
As a matter of fact,
the first curve describes an attractor of all the `neighboring'
solutions while the second one
`repels' them.

\begin{figure}
\includegraphics{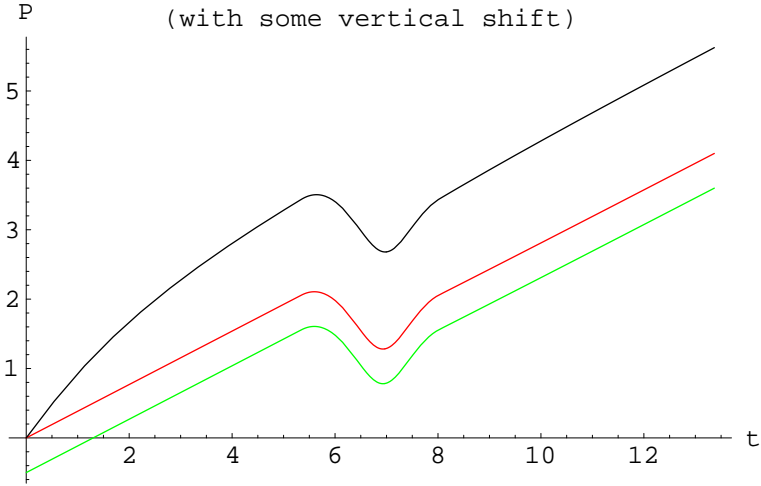}
\caption{\small
$P$-integrals corresponding to the stable steady \phaselocking{} are shown.
The black curve is the $P$-integral for the ground phase function
$\phi_0$ (the black curve in Fig.\  \ref{f12}).
The red curve represents $P$-integral is associated with
the refined \phaselocking{} candidate
determined by means of Eq.\ (\ref{eq181}).
The green curve represent the same function obtained
directly from definition
(\ref{eq4}), afterwards
it being 
shifted downward by 0.5 units.
}
\label{f15}
\end{figure}

\begin{figure}
\scalebox{0.6}[0.6]{
\includegraphics{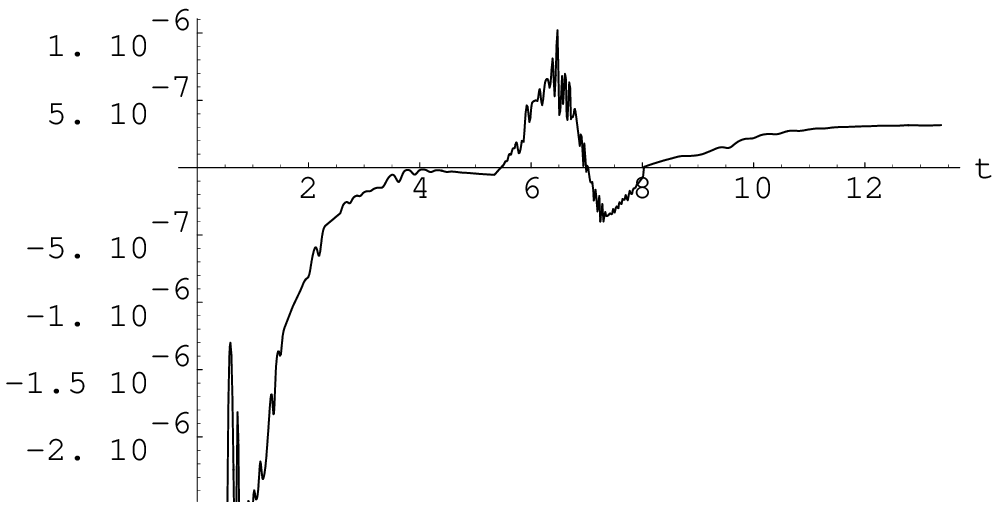}}
\scalebox{0.6}[0.6]{
\includegraphics{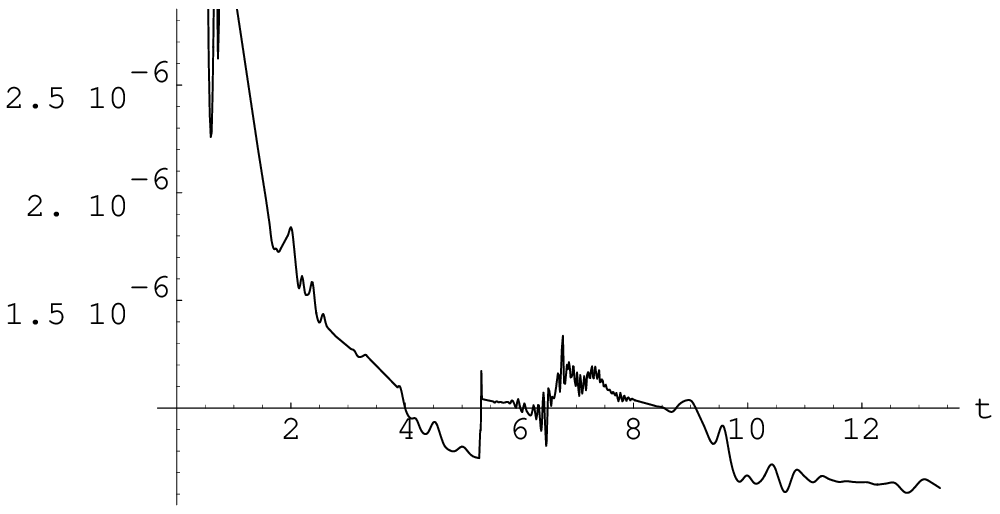}}
\caption{\small
The relative discrepancies of the different ways of computation of
$P$- and $Q$-integrals (by means of Eq.\ (\ref{eq181}) and the computation
in accordance with definitions from the known phase
functions) corresponding to the stable \phaselocking{}.
}
\label{f18}
\end{figure}

Above,
we have numerically demonstrated the validity of the
equation (\ref{eq181}) allowing one to build a new solution from the
known one.
Similarly,
Eq.\ (\ref{eq14}) allows one to determine, knowing $C$,
the integrals $P({t})$ and $Q({t})$ associated with the former.
Figures \ref{f15} and \ref{f18}
displays  the results of the corresponding computation.
There
the black curves represent the $P$- and $Q$-integrals
corresponding to the `ground' phase function
$\phi_0$ which is represented by black graph in Fig.\  \ref{f12}.
The red curves are the integrals for the refined \phaselocking{} candidates
which are determined by means of Eq.\ (\ref{eq181}).
The green curves (which are shifted in vertical direction downward) are
the same functions but they are obtained
by straightforward numerical integrating in accordance with
$P$ and $Q$ definitions. In Fig.\ \ref{f18},
the relative discrepancies of the two ways
of computation  of the $P$- and $Q$- integrals
are displayed. (Note that the relative discrepancy growth observed 
near the left boundary ${t}=0$
is the numerical artifact
caused by the vanishing of $P(0)$ and $Q(0)$.)

\begin{figure}
\includegraphics{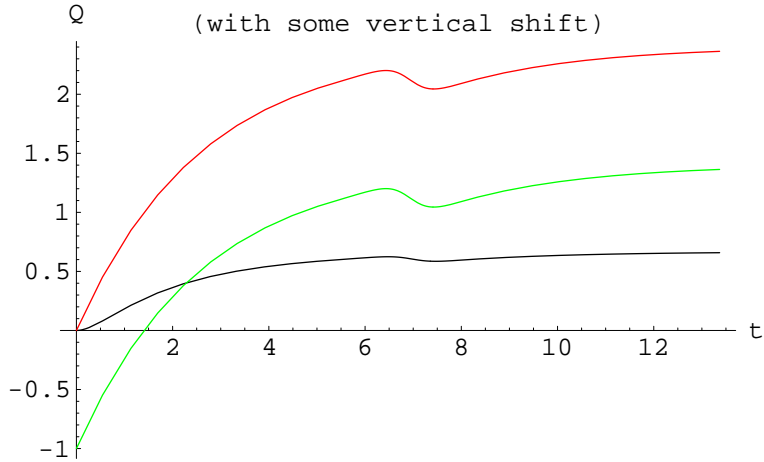}
\caption{\small
$Q$-integrals corresponding to the same stable \phaselocking{} state
as in Fig.\ \ref{f15} are shown. The color meaning is the same as therein.
The green graph is shifted downward by 1 unit.}
\label{f17}
\end{figure}

\begin{figure}
\includegraphics{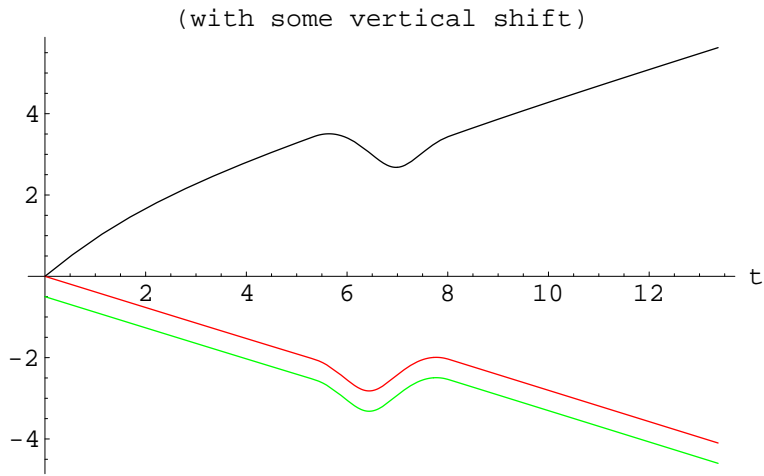}
\caption{\small
The graphical matter
displayed is
similar to the one shown in
Fig.\ \ref{f15} 
but it is connected with the unstable steady phase evolution
whose phase function is represented by the upper red curve in Fig.\ \ref{f12}.
}
\label{f15a}
\end{figure}

\begin{figure}
\includegraphics{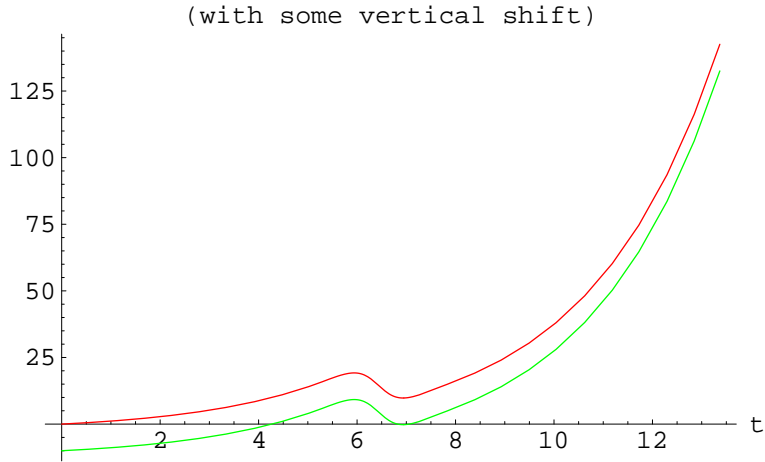}
\caption{\small
Similar to Fig.\ \ref{f17}
but
corresponds to the unstable \phaselocking{} candidate.
The `ground' $Q$-integral corresponding to the `ground' phase
function
(the analogue to the black graph in Fig.\ \ref{f15a})
is not shown because of the too large distinction in
the magnitudes.
}
\label{f17a}
\end{figure}

\begin{figure}
\scalebox{0.6}[0.6]{
\includegraphics{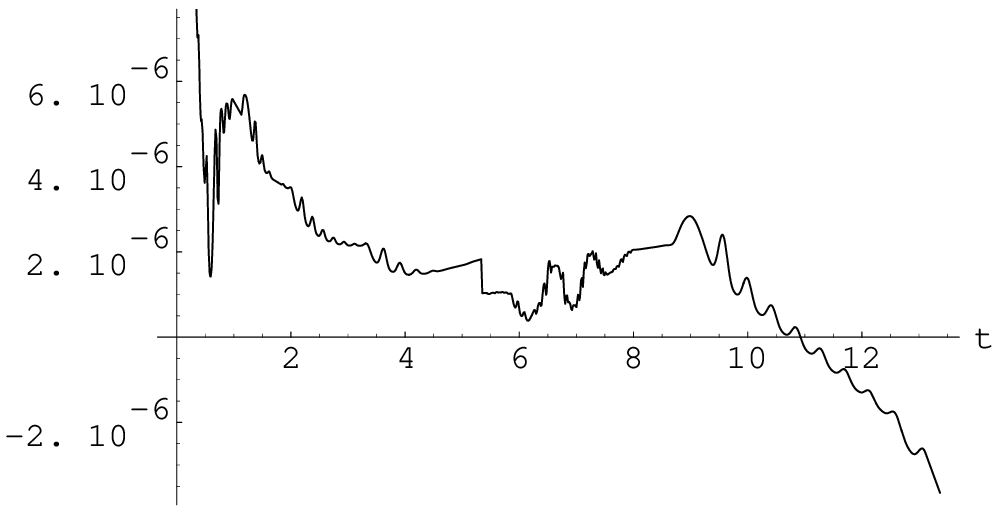}}
\scalebox{0.6}[0.6]{
\includegraphics{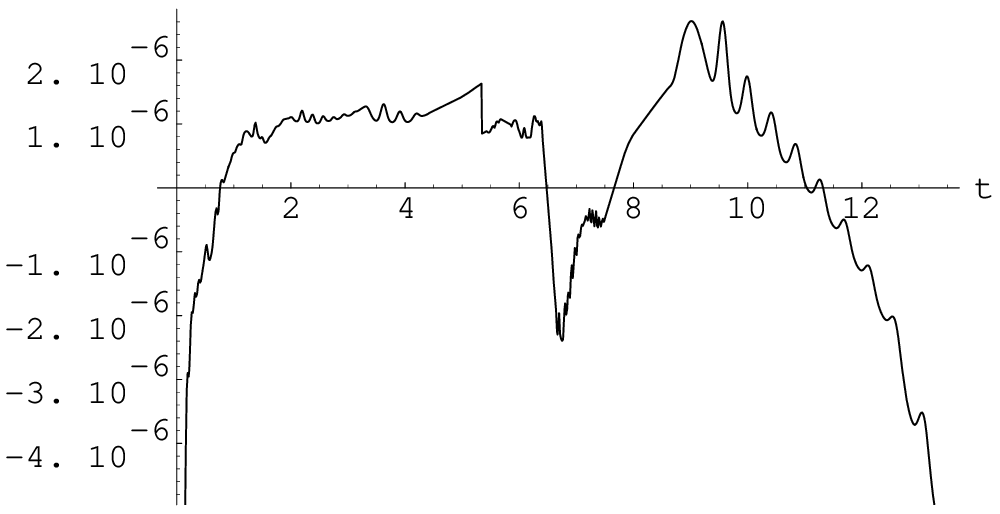}}
\caption{\small
The relative discrepancies of the results of two different ways of computation of
$P$- and $Q$-integrals corresponding to the unstable \phaselocking{}
candidate.
}
\label{f18a}
\end{figure}

\subsection*{$T$-scale discretization of the phase evolution}

As it has been noted, any  phase function $\phi({t})$
defined on arbitrary $t$ domain
can be equivalently described by the sequence of the functions
$\phi_j({t})$ defined on
the common interval
$[0,2\pi]$, see Eq.\ (\ref{eq16}).
At the same time, as a solution of Eq.\ (\ref{eq1X}),
each function $\phi_j({t})$
can be represented in the form  (\ref{eq11})
for the own specific value of the constant $C$ which we denote $C_j$.
A simple but important observation reads:
\begin{quote}
The property of \phaselocking{}  equivalent to 
the existence of the limit 
$\phi_\infty({t})$  of the sequence of
{\em functions\/} $\phi_j({t})$ (see Eq.\ (\ref{eq17}))
associated with a generic solution of Eq.\ (\ref{eq1X}),
is also equivalent to the existence of the limit  $C_{\infty}$ of the
sequence of {\em real constants\/} $C_j$, connected with $\phi_j$, either finite or infinite.
\end{quote}
Thus
 \begin{equation}
C_\infty=\lim_{j\rightarrow\infty} C_j
                        \label{eq211}
 \end{equation}
must exist, either finite or infinite, provided the \phaselocking{} takes
place.

On the contrary, the \phaselocking{}  does  not arise  and the phase
function reveals apparently
{{\chao}} behavior
if and only if the sequence
$C_j$ has no limit.

\medskip
\noindent
{\it Remark:}
\begin{quote}
Allowing constants $C_j$ and their
limit to assume infinite values, 
the real projective line $RP^1$ isomorphic to the
circumference has to be adopted
as the space of their ($C$-constants) originating.
This can be realized
by means of identifying each
$C_j$ with the pair $(C_j,1)$ and the
interpreting  the latter as homogeneous coordinates on $RP^1$. 
The limit of the sequence $C_j$ has also to be understood as a point in $RP^1$.
\end{quote}

\begin{figure}
\includegraphics{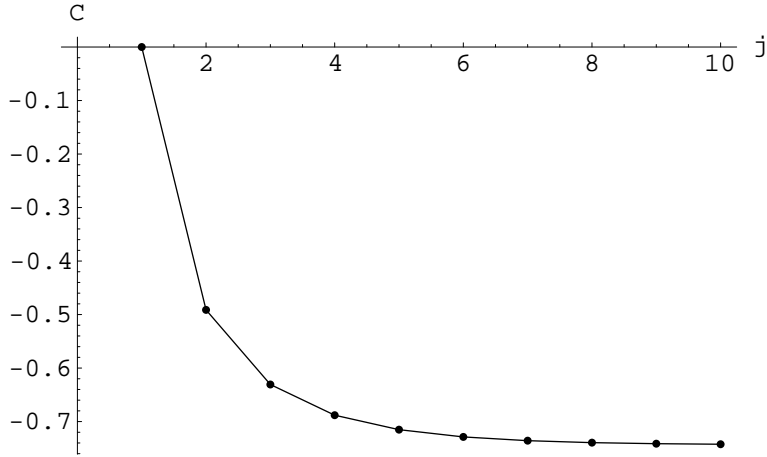}
\caption{\small The sequence of converging $C_j$ which
indicates the \phaselocking{} property is shown.
The bias is the same periodic rectangular pulse sequence as above besides
the specific value of the DC contribution
$\iota_{\mbox{\scriptsize dc}}=1.46$.
}
\label{f19}
\end{figure}

\begin{figure}
\includegraphics{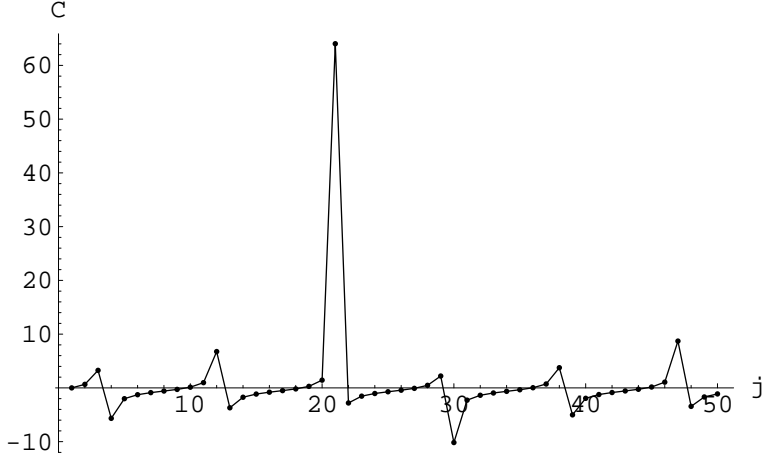}
\caption{\small Here the sequence $C_j$ not tending
to an apparent limit 
indicates
the absence of the \phaselocking{} ({{\chao}} phase behavior).
The bias is the same periodic rectangular pulse sequence as above besides
the specific value of the DC contribution
$\iota_{\mbox{\scriptsize dc}}=1.40$.
}
\label{f20}
\end{figure}

Figures \ref{f19}, \ref{f20} display the two examples of $C$-sequences for
the \phaselocking{} (Fig.\ \ref{f19}) and {{\chao}} (Fig.\ \ref{f20})
phase evolutions, respectively.
The distinction arises because  to the slightly different
values of the constant bias parameter:
$\iota_{\mbox{\scriptsize dc}}=1.46$ for Fig.\ \ref{f19}, and
$\iota_{\mbox{\scriptsize dc}}=1.4$ for Fig.\ \ref{f20}, 
cf.\ the left-lower panel of Fig.\ \ref{f11}.
The other parameters of the bias function $f$ are the same as
for the pulse
shown in the inset in
Fig.\ \ref{f6}.

In order to detect the \phaselocking{} without a plain
phase simulation,
one may,
in spite of determination of
the constants
$C_j$ from the functions
$\phi_j({t})$ by means of Eq.\ (\ref{eq80}),
make use of the following recurrence relation
which can be derived from
Eqs.\ (\ref{eq16}), (\ref{eq16a}), (\ref{eq13}) and $C_j$ definition:
{\small 
 \begin{equation}
C_{j+1}=
{
C_j
\left(e^{-\half P_{(0)}(T)}\cos\half\phi_{(0)}(T) -
Q_{(0)}(T)e^{\half P_{(0)}(T)}\sin\half\phi_{(0)}(T)\right)
-e^{\half P_{(0)}(T)}\sin\half\phi_{(0)}(T)
\over
C_j
\left(e^{-\half P_{(0)}(T)}\sin\half\phi_{(0)}(T)
+ Q_{(0)}(T)e^{\half P_{(0)}(T)}\cos\half\phi_{(0)}(T)\right)
+e^{\half P_{(0)}(T)}\cos\half\phi_{(0)}(T)
}
                         \label{eq24}
 \end{equation}
}
(It is worth reminding that  
the subscript `$_{(0)}$' hints at the specific initial conditions
the functions $\phi_{(0)},P_{(0)},Q_{(0)}$
obey
which read $  \phi_{(0)}(0)=0, P_{(0)}(0)=0, Q_{(0)}(0)=0.$
It 
emphasizes  the distinction of the `ground'
solution and its derivates from the particular element $\phi_0$ of the sequence
$\phi_j$; we shall omit this notation complication whenever possible.)
The relation above is of a notable importance allowing one
to obtain, finally, the exhaustive description
of the long-term phase behavior on the time scales exceeding $T$
by means of pure algebraic manipulations.

To that end, one has to mention that
the map $C_j \rightarrow C_{j+1}$ implied by Eq.\ (\ref{eq24}) is fraction linear.
It is well known that any such transformation
is equivalent
to a linear automorphism on a projective space.
To make use of much simplification
following from such a linear reduction
of the problem%
,
let us consider the vector space of 2-element columns
$\mathbf{C}=\left[{a \atop b}\right]$
(using here and below boldface characters for notation of matrix-valued quantities)
with the ratio of the elements equal to $C$, $a/b=C$.
We consider the columns differing by a non-zero
multiplier as equivalent,
$\left[{a \atop b}\right] \sim \left[{A a \atop A b}\right], A\not=0$.
Let us also introduce the $2\times2$ matrix
{\small 
\begin{equation}
\mathbf{\Phi}=
\left(%
\begin{array}{cc}%
e^{-\half P_0(T)}\cos\half\phi_0(T) - Q_0(T)e^{\half P_0(T)}\sin\half\phi_0(T)
&
-e^{\half P_0(T)}\sin\half\phi_0(T)
\\
e^{-\half P_0(T)}\sin\half\phi_0(T) + Q_0(T)e^{\half P_0(T)}\cos\half\phi_0(T)\
&
+e^{\half P_0(T)}\cos\half\phi_0(T)
\end{array}
\right)
                         \label{eq25}
\end{equation}}\
which is, as a straightforward
check shows, unimodular, $\det \mathbf{\Phi}=1$.
Then it is also straightforward to show that the  transformation
 \begin{equation}
 \mathbf{C}_{j+1}=
 \mathbf{\Phi}
 \mathbf{C}_{j}.
                         \label{eq26}
\end{equation}
is equivalent
to Eq.\ (\ref{eq24}).
The map $\mathbf{C}_{j}\mapsto \mathbf{C}_{j+1}$ corresponding to Eq.\
(\ref{eq26}) is just the linear automorphism of
the real projective line $RP^1$ 
mentioned above.

Since the transformation associated with matrix $\mathbf{\Phi}$ does not depends on $j$, 
starting from $j=0$, after $j-1$ iterations one  comes
to the equation
 \begin{equation}
 \mathbf{C}_{j}=
 \mathbf{\Phi}^j
 \mathbf{C}_{0}.
                         \label{eq27}
\end{equation}
It is convenient to chose
$\mathbf{C}_{0}=\left[{C_0 \atop 1}\right]$.
The very
`starting'
$C$-constant $C_0$ 
encodes the initial value of the phase function $\phi(t)$.
Giving $\phi(0)$, it can be calculated by means of
Eq.\ (\ref{eq80}).

Eq.\ (\ref{eq27}) is, essentially, the desirable algebraic
relationship describing
the long-term phase evolution.
The time variable $t$ is here encoded
in the integer variable $j$ playing role of a discrete time on the scale $T$
and amounting, numerically, to
the integer part of $t/T$.

We shall derive below the expanded version of (\ref{eq27})
where the power of the
operator $\mathbf{\Phi}$ is given in explicit form.

\subsection*{ The {\chao}{}
  phase evolution}

We consider,  first, the case
of {{\chao}} phase evolution when the discriminant $\Delta$, 
Eq.\ (\ref{eq20}), is negative.
In this case we proceed with the problem of calculation of
the accumulated phase variation over a long time interval
(the single period averaging does not yield a meaningful result here).
In accordance with (\ref{volt}),
it immediately
yields us the value of
the average voltage across the junction,
i.e.\ the  physically measurable
quantity.
In particular,
we shall see
that, in spite of apparently
{{\chao}} phase behavior, the 
phase evolution 
manifests actually no signs of chaos.
Moreover,
as a matter of fact, 
the phase
reveals, in a sense,
oscillation of a definite
frequency and the apparent irregularity
of its evolution
is manifested
because this frequency is in general incommensurable
with the bias guiding one (quasiperiodic behavior \cite{{X5}}).
As a particular consequence, 
the average voltage 
converges to a quite definite value
which can be calculated in a general case.

Elaborating the relationships referred to above,
let us
calculate, at first, the eigenvalues
of the matrix $\mathbf{\Phi}$.
In view of its role in the phase evolution,
one should not be surprised that the
discriminant of its characteristic equation  
 \begin{equation}
\lambda^2+1
-2\lambda
\left(
\cosh\half P_0(T) \cos\half\phi_0\left(T\right)
-\half e^{\half P_0(T)}
Q_0\left(T\right)\sin\half\phi_0\left(T\right)
\right)=0
                                                \label{eq28}
 \end{equation}
coincides,
up to the positive factor $4e^{-P_0(T)}$, with $\Delta\equiv\Delta[f]$,
see (\ref{eq20a}).
Thus, it is precisely the case of the {{\chao}} phase evolution when the matrix
(\ref{eq25}) has a pair of complex (and complex
conjugated) roots $\lambda_\pm$.
Furthermore, since their product is the unity,
they equal
$\exp\pm\imi\alpha$ for some real 
$\alpha$.
More concretely, one gets 
 \begin{eqnarray}
\lambda_\pm&=&e^{\pm \imi \alpha}=
                                        \label{eq28a}\\
&&\half
e^{\half P_0(T)}
\left(-Q_0(T)          \sin\half \phi_0(T)
      +(1+e^{-P_0(T)}) \cos\half\phi_0(T)
\vphantom{\sqrt{-\Delta}}
\right.
                                        \nonumber\\
&&
\hphantom{-\half e^{\half P_0(T)}}\left.\vphantom{e^{\half P_0(T)}}
     \pm \imi \sqrt{-\Delta}
\right).
 \end{eqnarray}
These eigenvalues are connected with the following
(complex valued) eigenvectors:
 \begin{equation}\label{eq28c}
\mathbf{V}_\pm=
\left[
\begin{array}{c}
- 2\sin\half\phi_0(T)         \\
Q_0(T) \sin\half \phi_0(T)
+(1-e^{-P_0(T)}) \cos\half\phi_0(T)
 \pm \imi \sqrt{-\Delta}
\end{array}
\right].
  \end{equation}
In other words, 
$$
\mathbf{\Phi}\mathbf{V}_\pm=e^{\pm \imi \alpha}\mathbf{V}_\pm.
$$

Further, the following identical decomposition of an arbitrary
real-valued 2-element column $\mathbf{A}$
in terms of the columns $\mathbf{V}_\pm$ takes place:
 \begin{equation}\label{eq:otyrfdl}
{\mathbf{A}}\equiv
\left(
\begin{array}{c}
A\\{} B
\end{array}
\right)
=K_+({\mathbf{A}})\,\mathbf{V}_++K_-({\mathbf{A}})\,\mathbf{V}_-
=2\Re(K_+({\mathbf{A}})\,\mathbf{V}_+),
  \end{equation}
where we make use of  the notation
 \begin{eqnarray}
 K_\pm({\mathbf{A}})&=&
\pm \imi
\left[4 \sin\half\phi_0(T) \sqrt{-\Delta}\right]^{-1}\times
                                \nonumber\\
&&
\left[
A\left(Q_0(T)\sin\half\phi_0(T) +(1-e^{-P_0(T)})\cos\half\phi_0(T)
\mp \imi\sqrt{-\Delta}\right)
\right.
                                \nonumber\\
&&
\left.
+ 2 B \sin\half\phi_0(T)
\vphantom{\sqrt{-\Delta}}
\right].
  \end{eqnarray}
The $j$-fold ($j=0,1,2,\dots$)
application of the linear matrix operator $\mathbf{\Phi}$ 
to the vector $\mathbf{A}$ expanded in accordance with (\ref{eq:otyrfdl}) leads to the equation
\begin{equation}
\mathbf{\Phi}^j {\mathbf{A}}
=
e^{\imi j\alpha}K_+({\mathbf{A}})\,\mathbf{V}_+
+
e^{-\imi j\alpha}K_-({\mathbf{A}})\,\mathbf{V}_-.
\label{eq:nfhtrse}
\end{equation}
This
is in fact the desirable explicit representation of the operator power
$\mathbf{\Phi}^j$ convenient for our purposes.

Applying the decomposition (\ref{eq:nfhtrse}) to Eq.\ (\ref{eq27}),
the resulting equation can then be resolved 
with respect to $C_j$ representing it as a fraction-linear function of
$C_0$, 
Its formula will be given later on (see Eq. (\ref{eq42}))
while here we
write down instead the explicit form of the equation
\begin{equation}
e^{\imi\phi(j T)}=e^{\imi\phi_j(0)}=e^{\imi\phi_{(0)}(0)}
{1+ C_j \overline{{\cal F}_{(0)}(0)}
\over 
1+ C_j{\cal F}_{(0)}(0)
}={1-\imi C_j\over 1+\imi C_j}
\end{equation}
following from the definition of  $C_j$
(and making use of the initial values $\phi_{(0)}(0)=0,{\cal
  F}_{(0)}(0)=i$,
following, in turn, from definitions of $\phi_{(0)},{\cal F}_{(0)} $).
It reads
 \begin{equation}
e^{\imi\phi(j T)}=
{
e^{\imi j\alpha} U^{(+)} +e^{-\imi j\alpha} U^{(-)}
\over
e^{-\imi j\alpha} \overline{U^{(+)}}
+e^{\imi j\alpha} \overline{U^{(-)}}
}%
,\enskip j=0,1,2,\dots,
\label{eq:uyrvhyd}
 \end{equation}
where all the coefficients
 \begin{eqnarray}\label{eq:jfyrser}
U^{(+)}&=&n_R +\sqrt{-\Delta}
 -\imi\left(n_I+C_0\sqrt{-\Delta}\right),
                                            \nonumber \\
U^{(-)}&=&-n_R +\sqrt{-\Delta}
 +\imi\left(n_I-C_0\sqrt{-\Delta}\right),
                                            \nonumber \\
n_R&=&
C_0(1-e^{-P_{(0)}(T)})\cos\half\phi_{(0)}(T)+(C_0 Q_{(0)}(T)+2)\sin\half\phi_{(0)}(T),
                                            \nonumber \\
n_I&=&
(1-e^{-P_{(0)}(T)}+2C_0 Q_{(0)}(T) )\cos\half\phi_{(0)}(T)  
                                          \nonumber \\
&&
+\left(2 C_0 e^{-P_{(0)}(T)}+ Q_{(0)}(T) \right)\sin\half\phi_{(0)}(T)
\end{eqnarray}
do not depend on $j$ and $n_R,n_I$ are real. (Notice 
that the only `complex valued
ingredient' in  $U^{(\pm)}$ is the factor $\imi$.)
Since $j$ represent  here the time
`normalized and discretizied on the
scale $T$'
(i.e.\ the integer part of $t/T$, in fact),
Eq.\ (\ref{eq:uyrvhyd})
manifest,  essentially,
the specific `hidden' periodicity of the phase function 
which has been occasionally
referred to above, the corresponding period amounting to $2\pi
T\alpha^{-1}$.
Generally speaking, the latter quantity
is incommensurable with the bias period
$T$.
As a consequence, the discrete parameter $j$ never assumes 
sequential 
values differed
by  $2\pi T\alpha^{-1}$ or a quantity aliquot to it.
It is this circumstance which
does not allows one to distinguish the oscillations
which would be described by Eq.\ (\ref{eq:uyrvhyd}) if $j$ were a
continuous variable.

\subsection*{Average rate of the phase growth}

Now
we are in  position to
apply the relationships
derived above for the 
determination of the {\em average voltage\/} ${V}_{\mbox{\scriptsize av}}$
across junction in the
case of {{\chao}} phase behavior (\phaselocking{} absence).
In view of Eqs.\ (\ref{eq2}), (\ref{volt}),
 \begin{equation}
 {2\pi T\over \Phi_0}{V}_{\mbox{\scriptsize av}}^{(j)}=
{1\over j}[\phi(j T)-\phi(0)]
 \end{equation}
and, therefore.
 \begin{eqnarray}
\exp
 {2 \imi \pi T\over \Phi_0}{V}_{\mbox{\tiny av}}^{(j)}
&=&
\left\{
e^{\imi\phi(j T)}e^{-\imi\phi(0)}
\right\}^{1\over j}
                   \nonumber\\
&=&
\left\{
{1+\imi C_0
\over
 1-\imi C_0
}
\right\}^{1\over j}
%
%
\left\{
{
e^{\imi j\alpha} U^{(+)} +e^{-\imi j\alpha} U^{(-)}
\over
e^{-\imi j \alpha} \overline{U^{(+)}}
+e^{\imi j \alpha} \overline{U^{(-)}}
}%
\right\}^{1\over j}.
                                \label{eq29}
 \end{eqnarray}

Thus the averaging over a lapse 
consisting of unboundedly increasing
number of bias periods reduces to the
computation of the limit
${V}_{\mbox{\tiny av}}^{(\infty)}=\lim_{j\rightarrow\infty}
{V}_{\mbox{\tiny av}}^{(j)}$ for ${V}_{\mbox{\tiny av}}^{(j)} $
satisfying (\ref{eq29}).
It yields a definite result if and only if
the latter exists. 
Let us consider how it can be computed.

As to the first multiplier involved
in Eq.\ (\ref{eq29}), its nonzero base does not
depend on $j$ and the limit
always exists
(obviously, it equals the unity).
On the contrary,
the second multiplier is not a universal constant.
It 
depends on the ratio of the modules of the
complex coefficients
$U^{(+)}$ and $U^{(-)}$.

Clarifying the latter point, let us
calculate the difference
$\Sigma=\vert U^{(+)}\vert^2-\vert U^{(-)}\vert^2$.
A straightforward algebra yields
\begin{eqnarray}
\Sigma&=&8\sqrt{-\Delta}e^{P_{(0)}}\Sigma_0,\enskip\mbox{where}
\nonumber\\
\Sigma_0&=&
C_0^2\left(e^{P_{(0)}}Q_{(0)}\cos{\half\phi_{(0)}}+\sin{\half\phi_{(0)}}\right)
\nonumber\\&&
+
C_0\left(\left(e^{P_{(0)}}-1\right)\cos{\half\phi_{(0)}}+e^{P_{(0)}}Q_{(0)}\sin{\half\phi_{(0)}}\right)
+e^{P_{(0)}}\sin{\half\phi_{(0)}}
\end{eqnarray}
(the argument $(T)$ is here omitted, cf.\ (\ref{eq:jfyrser})).
The sign of the difference coincides therefore with the sign of the factor $\Sigma_0$,
Apart from the constant parameters determining
$\phi_{(0)},P_{(0)},Q_{(0)} $,  $\Sigma_0$ also
depends on $C_0$ which encodes the starting (initial) value of
the phase
and may assume, in principle, arbitrary real value including
the limiting
case $C\rightarrow\infty$.
However, $\Sigma_0$ is itself a second order polynomial in $C$.
Calculating its discriminant,  one finds it to be 
connected, again, with $\Delta[f]$,
equaling to
\begin{equation}
 e^{2 P_0(T)}\Delta,
                                \label{eq31}
\end{equation}
and being in our case {\em negative\/}
just in view of the {{\chao}} phase behavior condition $\Delta<0$.
Therefore, as a function of $C_0$, $\Sigma_0$ has no real roots 
and is either everywhere  positive or everywhere negative
irrespectively of the initial phase (connected with the
specific value of $C_0$).
Furthermore, since
\begin{equation}
\Sigma_0\vert_{C_0=0}=
e^{P_0(T)} \sin{\half\phi_0(T)},
                                \label{eq32}
\end{equation}
the sign of $\Sigma_0$ (and $\Sigma$) 
coincides with the sign of  $\sin{\half\phi_0(T)}$
which is determined, ultimately, by the function $f$ and
may not vanish by virtue of definition of $\Delta$
and the `no convergence' condition $\Delta<0$, see (\ref{eq20}).

Thus we have shown that
\begin{quote}
$\vert U^{(+)}\vert> \vert U^{(-)}\vert$ iff
$\sin{\half\phi_0(T)}>0$\\ and\\
$\vert U^{(+)}\vert< \vert U^{(-)}\vert$ iff
$\sin{\half\phi_0(T)}<0$.
\end{quote}

One may employ any of the 
two decompositions shown below of the second multiplier
in the last line of (\ref{eq29})
 \begin{eqnarray}
\rlap{\mbox{$\displaystyle
\left\{
{
e^{\hphantom{-}\imi j\alpha} U^{(+)} -e^{-\imi j\alpha} U^{(-)}
\over
e^{-\imi j \alpha} \overline{U^{(+)}}
-e^{\hphantom{-}\imi j\alpha} \overline{U^{(-)}}
}%
\right\}^{1\over N}
     $}}
                        \nonumber\\
&\equiv&
\left\{
\begin{array}{l}
\displaystyle
e^{\hphantom{-}
2\imi \alpha}
\left({U^{(+)} \over\overline{ U^{(+)}}}\right)^{1\over j}
\left\{
{
1 -e^{-2 \imi j\alpha} \left(U^{(-)}/U^{(+)}\right)
\over
1 -e^{\hphantom{-} 2 \imi j\alpha} \overline{\left(U^{(-)}/U^{(+)}\right)}
}%
\right\}^{1\over j}
\\
\displaystyle
e^{-2\imi \alpha}
\left({U^{(-)} \over\overline{ U^{(-)}}}\right)^{1\over j}
\left\{
{
1 -e^{\hphantom{-}2 \imi j\alpha} \left(U^{(+)}/U^{(-)}\right)
\over
1 -e^{-2 \imi j\alpha} \overline{\left(U^{(+)}/U^{(-)}\right)}
}%
\right\}^{1\over j}
\end{array}
\right.
                                \label{eq33}
 \end{eqnarray}
Let us arrange to
apply the upper representation if
$\vert U^{(+)}\vert> \vert U^{(-)}\vert$ and the lower one
in the opposite case. 
Then the limit of the
last multiplier in the corresponding line
of (\ref{eq33}) exists and are equal to the unity
since its base to be raised to the power $1/j$ is a finite $j$-independent gap
distant from zero.
The existence
of limits of the second multiplier is evident,  it also equals the unity.
Thus the limit of the first (constant) factor yields
 the limit value for the whole product.

Having thus calculated the limit of (\ref{eq33}) as $j\rightarrow\infty$,
Eq.\ (\ref{eq29}) leads to the 
following conclusion:
\begin{eqnarray}
\exp{
 {2 \imi \pi T\over \Phi_0}V_{\mbox{\tiny av}}^{(\infty)}
} &=&
\left\{
\begin{array}{l}
\displaystyle
e^{2\imi \alpha} \enskip\mbox{for}\enskip \sin{\half\phi_0(T)}>0
\\
\displaystyle
e^{-2\imi \alpha}\enskip\mbox{for}\enskip \sin{\half\phi_0(T)}<0
\end{array}
\right.
\end{eqnarray}
where $\alpha$ is defined in Eq.\  (\ref{eq28a}).
This simple result yields 
the explicit representation of
the average voltage across junction in the case
of {{\chao}} phase evolution:
\begin{eqnarray}
V_{\mbox{\tiny av}}^{(\infty)}
 &=&
{\Phi_0 \over \pi T}\times
\left\{
\begin{array}{l}
\displaystyle
 \alpha+k\pi \enskip\mbox{if}\enskip \sin{\half\phi_0(T)}>0
\\
\displaystyle
- \alpha+k\pi\enskip\mbox{if}\enskip \sin{\half\phi_0(T)}<0
\end{array}
\right.
                                \label{eq34}
\end{eqnarray}
for some integer $k$ 
(we shall discuss the method of its determination later on).

More exactly, the formula above 
describes, for a single $k$, a single branch
binding the
two neighboring Shapiro steps, \ie{}
`horizontal' constant voltage segments
on the junction I-V curve whose orders differ by the unity.
Such branches are sometimes called `resistive portions of I-V curve' \cite{X5}
although the specific dependence of the average voltage on  
$\iota_{\mbox{\scriptsize dc}}$,
as it is described by Eq.\ (\ref{eq34}),
may considerably deviate 
from the simple Ohm's law proportionality. In particular,
near the edges of these `resistive portions' the {\it differential
resistance\/} $\mathrm{d} V_{\mbox{\tiny av}}^{(\infty)}/ \mathrm{d} \idc$ diverges.

\subsection*{Solution of the spectral problem in the case $\Delta>0$ }

After a minor modification, the majority of
the above equations  is also
applicable 
in the case of \phaselocking{}.
Here one has to assume
 \begin{equation}
 \Delta>0.
 \end{equation}
Then Eqs.\ (\ref{eq24}-\ref{eq27}) hold  true
while the adapted versions
of Eqs.\ (\ref{eq28a})-(\ref{eq28c}) read
 \begin{eqnarray}
\lambda_\pm&=&
\half
e^{\half P_0(T)}
\left(-Q_0(T)          \sin\half \phi_0(T)
      +(1+e^{-P_0(T)}) \cos\half\phi_0(T)
\vphantom{\sqrt{\Delta}}
\right.
                                        \nonumber\\
&&
\hphantom{-\half e^{\half P_0(T)}}\left.\vphantom{e^{\half P_0(T)}}
     \pm \sqrt{\Delta}
\right),                                
                                \label{eq35x}
 \end{eqnarray}
 \begin{equation}
\mathbf{V}_\pm=
\left[
\begin{array}{c}
- 2\sin\half\phi_0(T)         \\
Q_0(T) \sin\half \phi_0(T)
+(1-e^{-P_0(T)}) \cos\half\phi_0(T)
 \pm \sqrt{\Delta}
\end{array}
\right].
  \end{equation}
These are the eigenvalues and the eigenvectors of the matrix
$\mathbf{\Phi}$  (\ref{eq25}), and the following
equations take place
$$
\mathbf{\Phi}\mathbf{V}_\pm= \lambda_\pm \mathbf{V}_\pm,
$$
where in this case $\lambda_\pm, \mathbf{V}_\pm$ are real.
It is worth noting that the unimodularity constraint
 \begin{equation}
\lambda_+\cdot \lambda_-=1
                                \label{eq38}
 \end{equation}
also holds true.
Then it follows from Eq.\ (\ref{eq27}) 
 \begin{equation}
-4 \sin\half\phi_0(T) \sqrt{\Delta}
\mathbf{C}_{j}
=\lambda_+^j\tilde K_+\,\mathbf{V}_+ + \lambda_-^j\tilde K_-\,\mathbf{V}_-,
                                \label{eq39}
  \end{equation}
where
 \begin{eqnarray}
\tilde K_\pm&=&\tilde K_\pm(C_0)=
C_0 L_\mp
+ 2 \sin\half\phi_0(T)
\vphantom{\sqrt{\Delta}},
                                \label{eq40}
                                \\
L_\pm&=&
Q_0(T)\sin\half\phi_{0}(T) +(1-e^{-P_0(T)})\cos\half\phi_{0}(T)
\pm \sqrt{\Delta},
                                \label{eq41}
  \end{eqnarray}
and, after some algebra, one gets
 \begin{eqnarray}
C_j&=&
- 2\sin\half\phi_{0}(T)
{
\lambda_+^j\tilde K_+(C_0)+ \lambda_-^j\tilde K_-(C_0)
\over
\lambda_+^j\tilde K_+(C_0)L_+ + \lambda_-^j\tilde K_-(C_0)L_-
}.
                                \label{eq42}
\\&=&
{-2\sin\half\phi_{0}(T)}\times
\nonumber\\&&
{ (L_- C_0 +2\sin\half\phi_{0}(T))\lambda^{j}_+
- (L_+ C_0 +2\sin\half\phi_{0}(T))\lambda^{j}_-
 \over
(L_- C_0 +2\sin\half\phi_{0}(T))L_+\lambda^{j}_+
- (L_+ C_0+2\sin\half\phi_{0}(T))L_-\lambda^{j}_-
  }.\nonumber
  \end{eqnarray}

Here all the dependence on  $j$ 
(proportional, up to normalization and discretization, to the
evolution time $t$)
is isolated in the factors $\lambda_\pm^j$. More precisely,
the  $j$ dependent terms combine
to the powers $(\lambda_+/\lambda_-)^j$ or
$(\lambda_-/\lambda_+)^j$ but, in view of
Eq.\ (\ref{eq38}), these coincide with $\lambda_\pm^{2j}$, respectively.

\subsection*{Calculation of phase function}

Basing on (\ref{eq42}), the explicit equation
{\em completely determining the phase\/} at any moment of time
in terms of the `ground' phase function 
$\phi_{(0)}$ specified on the interval  $[0,T]$
and the functions $P_{(0)},Q_{(0)}$ calculated  from $\phi_{(0)}$
and defined on the same interval
is easily derived:
\begin{eqnarray}
e^{\imi\phi(t'+jT)}&=&
e^{\imi\phi_{(0)}(t')}\times
\nonumber\\&&\hspace{-3em}
{
(M_+L_+\lambda^{j}_+
 -
 M_-L_-\lambda^{j}_-)
 -
2\sin\half\phi_{(0)}(T)
 (M_+\lambda^{j}_+
 -
 M_-\lambda^{j}_-
 )\overline{{\cal F}_{(0)}(t')}
 \over
(M_+L_+\lambda^{j}_+
 -
 M_-L_-\lambda^{j}_-)
 -
2\sin\half\phi_{(0)}(T)
 (M_+\lambda^{j}_+
 -
 M_-\lambda^{j}_-
 ){{\cal F}_{(0)}(t')}
},
                    \label{eq533} \\
&=&
e^{\imi\phi_{(0)}(t')}\times \nonumber
                    \\&&\hspace{-6em}
{
(L_+-2\sin\half\phi_{(0)}(T)\overline{{\cal F}_{(0)}(t')})M_+\lambda^{j}_+
-
(L_--2\sin\half\phi_{(0)}(T)\overline{{\cal F}_{(0)}(t')})M_-\lambda^{j}_-
\over
(L_+-2\sin\half\phi_{(0)}(T){{\cal F}_{(0)}(t')})M_+\lambda^{j}_+
-
(L_--2\sin\half\phi_{(0)}(T){{\cal F}_{(0)}(t')})M_-\lambda^{j}_-
},
\label{eq534}
\end{eqnarray}
where
\begin{eqnarray}
M_\pm&\equiv&M_\pm(C_0)=L_\mp C_0 +2\sin\half\phi_{(0)}(T),
%
\label{eq534a}
\end{eqnarray}
$t'\in[0,T)$ is the excess of $t$ over the nearest lower $j T$,
$t'=t-T[[t/T]]$,
for integer $j=[[t/T]]$.

\medskip
{\it Remark:}
\begin{quote}
Eqs.\ (\ref{eq533})-(\ref{eq534a}) also apply to the case of {\chao}
phase evolution and exactly in this form.
The only difference is in the definitions of $\lambda_{\pm},L_{\pm}$ 
(\ref{eq35x}),(\ref{eq41}).
In {\chao}{} case one has to replace the real term $\sqrt\Delta$
by the pure imaginary $\imi\sqrt{-\Delta}$. Then
$\overline{\{\lambda_{\pm},L_{\pm},M_{\pm}\}}
=\{\lambda_{\mp},L_{\mp},M_{\mp}\}$ 
that, in particular, ensure the phase function defined by
  (\ref{eq534}) to be
real.
\end{quote}

Eq.\ (\ref{eq533})  determines
the phase  at any moment of time (up to a constant
summand aliquot to $2\pi$ which will be computed later on)
through the phase initial value
 $\phi(0)$,
the solution  $\phi_{0}$ of Eq.\ (\ref{eq1X}) vanishing at $t=0$
and the complex valued function ${\cal F}_{0}(t)$
both computed on the finite interval $[0,T]$.
The exhaustive description
of the process of the asymptotic establishing of the 
steady \phaselocking{} state is its particular byproduct.
It allows one to explicitly
describe the `transient processes'
representing the distinction of
arbitrary given phase function
from the `refined' steady one to which the former asymptotically converges.

To be more specific,
let us now consider in brief some straightforward
consequences of Eqs.\ (\ref{eq42}),(\ref{eq533}).

At first, it should be noted, that the dependence of r.h.s.\ of
(\ref{eq533})
on $j$ (the time normalized to the scale
$T$ and then `discretizied')
disappears in the case of satisfaction of any of the two equations
\begin{eqnarray}
 M_+=0\Leftrightarrow C_0&=&-2\sin\half\phi_{0}(T)/L_-\;\mbox{or}
                                     \label{eq534b}
\\
 M_-=0\Leftrightarrow C_0&=&-2\sin\half\phi_{0}(T)/L_+.
                                     \label{eq534c}
\end{eqnarray}
(It is worth noting that, requiring $C_0$ to be real,
each of them implies the condition $\Delta\ge 0$.
Thus, neither of Eqs. (\ref{eq534b}), (\ref{eq534c}) can be fulfilled
in the {\chao} case.)
Then $e^{\imi\phi}$ is a periodic function of $t$ with the period $T$.
In particular, it
coincides with the own asymptotic
limiting form.
Such phase evolution can be named 
a {\it
  steady\/} one.
Notice that the `steadiness' does not means the periodicity of $\phi(t)$
but allows, apart from periodic part, the uniformly growing contribution $2\pi k t/T$
  for an
integer $k$.

It can be shown by a direct check that $C_0$, satisfying
(\ref{eq534b}) or (\ref{eq534c}), also verifies Eq.\ (\ref{eq19}).
Hence, (\ref{eq534b}) and (\ref{eq534c}) determine
the $C$-constants
which correspond to initial data characteristic
of the `refined' (steady) \phaselocking{} states.
Two choices $\pm$ correspond to two states,
stable and unstable.
Such $C$-constants were referred to above as $C_\infty$.
Now let us arrange to reserve this notation for the $C$-constant
implying the {\em stable\/} steady phase evolution.

In order to determine which
of these two states is stable, let us
consider a generic case when both  $M_+\ne0\ne M_-$.
It follows from
(\ref{eq35x}) and (\ref{eq38}) that the condition
$\Delta>0$ implies either  $|\lambda_+|>1$ and $|\lambda_-|<1$
or $|\lambda_+|<1$ and $|\lambda_-|>1$ (and both $\lambda_\pm $ are either positive
or negative).
\begin{quote}\em
Let us arrange about the following interpretation of the subscripts `$max$'
and `$min$':
\begin{eqnarray}
  \label{eq:rtyet}
\mbox{if}\;  |\lambda_+|>1 \; \& \;|\lambda_-|<1&\mbox{then}
&
\mbox{`$max$' means `$+$', `$min$' means `$-$'},
\nonumber \\
\\
\mbox{if}\;  |\lambda_-|>1 \; \& \; |\lambda_+|<1&\mbox{then}
&
\mbox{`$max$' means `$-$', `$min$' means `$+$'}.
\nonumber
\end{eqnarray}
\end{quote}
Then one may introduce the equation
\begin{equation}
  e^{-2\varkappa}\equiv
  {\lambda_{min}\over\lambda_{max}}=  \left|
{
\left|-Q_{0}(T)\sin\half\phi_{0}(T)+(1+e^{-P_{0}(T)}) \cos\half\phi_{0}(T)\right|
  -\sqrt\Delta
\over
\left|-Q_{0}(T)\sin\half\phi_{0}(T)+(1+e^{-P_{0}(T)}) \cos\half\phi_{0}(T)\right|
  +\sqrt\Delta
} \right|<1,
\end{equation}
defining by them the {\em positive} constant $\varkappa $.
Using it,
Eq.\ (\ref{eq533}) can be recast  to the following form:
\begin{eqnarray}
e^{\imi\phi(t'+jT)}&=&
e^{\imi\phi_{0}(t')}\cdot
{ L_{max}-2\sin\half\phi_{0}(T)\overline{{{\cal F}_{0}(t')}}
\over
L_{max}-2\sin\half\phi_{0}(T){\cal F}_{0}(t')
}
\cdot
{1-\overline{{Z}(t')}e^{-2\varkappa j }
\over
1-{Z}(t')e^{-2\varkappa j}
},
                        \label{eq647}
                \\
\mbox{where }Z(t)&=&
{
L_{min}-2\sin\half\phi_{0}(T){\cal F}_{0}(t)
\over
L_{max}-2\sin\half\phi_{0}(T){\cal F}_{0}(t)
}\cdot
{M_{min}\over M_{max}}
                        \nonumber
\end{eqnarray}
Sending here $j\rightarrow\infty$,
one finds that the exponent $e^{\imi\phi(t)}$ for a `generic' solution%
\footnote{for which $Z(t)\not\equiv 0 $}
$\phi(t)$ of Eq.\ (\ref{eq1X})
approaches to the {\em periodic\/} function $e^{\imi\phi_\infty(t)}$
defined for $t\in[0,T]$ by the equation
\begin{equation}
e^{\imi\phi_\infty(t)}=
e^{\imi\phi_{0}(t)}\cdot
{ L_{max}-2\sin\half\phi_{0}(T)\overline{{\cal F}_{0}(t)}
\over
L_{max}-2\sin\half\phi_{0}(T){\cal F}_{0}(t)
}.
                                                \label{eq77}
\end{equation}
Notice that the mentioned periodicity is not manifest from this
formula,
it is rather a specific consequence of the properties of its ingredients 
constituting the essence of the definitions of the latter.

The very function of the `refined' \phaselocking{} steady phase
 evolution,
 $\phi_\infty(t)$,
advances at each time step of duration $T$ by $2\pi k$ for some
{\em fixed} ($t$-independent)
 integer $k$.
The formula (\ref{eq77}) is nothing else but Eq.\ (\ref{eq:ogytgfdr})
with
\begin{equation}
  C=C_\infty=
-{2\sin\half\phi_{0}(T)
\over
L_{max}}.
                                        \label{eq539}
\end{equation}
This answers the question which of the two equations
$M_+=0$ or $M_-=0$
describes the stable
steady evolution (asymptotic \phaselocking{} state):
it is described  by the equation
\begin{equation}
M_{min}=0.
\end{equation}

\subsection*{Phase-locking{} criterium}

The detailed picture
of the convergence of a generic phase function $\phi(t)$ to
the asymptotic limit
$\phi_\infty(t)$ can be inferred from the equation (\ref{eq647})
recast as follows:
\begin{eqnarray}
e^{\imi(\phi(t'+jT)-\phi_{\infty}(t'+jT))}&=&
{1-\overline{{Z}(t')}e^{-2\varkappa j}
\over
1-{Z}(t')e^{-2\varkappa j}
},\;t'\in[0,T),j=0,1,2,\dots
                        \label{eq649}
\end{eqnarray}
which describes what can be called a `nonlinear exponential' convergence.
If $j=[[t/T]]$ is sufficiently large to ensure the
satisfaction of the  condition
\begin{equation}
\max_{[0,T]}|Z|<e^{2\varkappa j},
\label{eqgvrsde}
\end{equation}
the simple estimate follows
$$|\phi(t'+jT)-\phi_{\infty}(t'+jT)|<\log(1+|Z|e^{-2\varkappa j})-\log(1-|Z|e^{-2\varkappa j})<
2\max|Z|e^{-2\varkappa j}
$$
which describes just the exponential convergence.
The factor $\max|Z|$ and the exponent coefficient $\varkappa$ both
determine how many periods (enumerated by the integer $j$)
have to  elapse until the phase  function
approaches the steady asymptotic state with the prescribed accuracy.

At the same time, for comparatively small time $t$, the `transients'
$\phi(t)-\phi_{\infty}(t)$ may have no relation to the exponent.
Eq.\ (\ref{eqgvrsde}) allows to estimate the duration of this
`near-zone'
lapse.

As to the second possibility
\begin{equation}
C=C_{\bowtie}\equiv-{2\sin\half\phi_{0}(T)\over L_{min}}
\end{equation}
 (the solution of the equation
$M_{max}=0$),
it also yields, formally,
the steady \phaselocking{} state described by the
phase function $\phi_{\bowtie}(t)$,
which can be computed with the help of equation
\begin{equation}
e^{\imi\phi_{\bowtie}(t)}=
e^{\imi\phi_{0}(t)}\cdot
{ L_{min}-2\sin\half\phi_{0}(T)\overline{{\cal F}_{0}(t)}
\over
L_{min}-2\sin\half\phi_{0}(T){\cal F}_{0}(t)
},
                                          \label{eq77a}
\end{equation}
and is distinct from $\phi_{\infty}(t)$.
This is unstable solution `repelling' any neighboring one.
Indeed, having rewritten Eq.\ (\ref{eq647}) as follows
\begin{eqnarray}
e^{\imi\phi(t+jT)}&=&
e^{\imi\phi_{\bowtie}(t)}
\cdot
{1-\overline{{\tilde Z}(t)}e^{2\varkappa j}
\over
1-{\tilde Z}(t)e^{2\varkappa j}
},
                        \label{eq648}
                \\
\mbox{where }\tilde Z(t)&=&
{
L_{max}-2\sin\half\phi_{0}(T){\cal F}_{0}(t)
\over
L_{min}-2\sin\half\phi_{0}(T){\cal F}_{0}(t)
}\cdot
{M_{max}\over M_{min}},
                        \nonumber
\end{eqnarray}
one sees that for arbitrary small but non-zero $M_{max}$
(which is non-zero if $\phi\not\equiv\phi_{\bowtie}$ mod $2\pi$)
the function
$e^{\imi\phi(t+jT)}$
escapes exponentially off the function
$e^{\imi\phi_{\bowtie}(t+jT)}$
which, itself,  can be revealed
only in the case of the exact vanishing of $M_{max}$.
For any other phase function 
$e^{\imi\phi(t+jT)}$  ultimately approaches, as $j$ increases, to $e^{\imi\phi_\infty}$.

\newtheorem{theo}{Theorem}

The computations above proves the following statement which has been
mentioned above:
\begin{theo}\label{gftdre}
 The condition $\Delta\ge0$ is necessary
 for the \phaselocking{} to be observed
 whereas the strict inequality $\Delta>0$ is sufficient.$\square$
 \end{theo}

{\em Remarks:}
\begin{itemize}\item There is  a formulation of the criterion above
operating with not specific but
arbitrary solution $\phi$ of Eq.\ (\ref{eq1X}) (instead of $\phi_{(0)}$)
and not requiring for the functional
$\cal F$ to obey the specific initial condition (\ref{eq86})%
\footnote{Nevertheless, $\Fn(0)$ may not vanish for any representation.}%
. In general form,
it reads
 \begin{eqnarray}
|D|&>& 1,\\
\mbox{where}\;
D&=&-
{
\Im[e^{-\halfi\Delta\phi}
(\Fn(t_0+T/2)-\overline{\Fn(t_0-T/2)})]
\over
2\sqrt{\Im[\Fn(t_0+T/2)]\Im[\Fn(t_0-T/2)]}
},
\label{eq:iroycd}
\\
\Delta\phi&=&\phi(t_0+T/2)-\phi(t_0-T/2).
\end{eqnarray}
$D$ does not depend on $t_0$.
\item
In terms of $D$, the eigenvalues $\lambda_{\pm}$ are represented as follows
$$\lambda_{\pm}=D\pm\sqrt{D^2-1}.$$
\item
It is shown below that for $\Delta=0$ the \phaselocking{}
is also observed but its properties reveal some distinction from
ones of 
the case $\Delta>0$ (the `weak' \phaselocking{} against the
`generic' one).
\end{itemize}

\subsection*{Weak \phaselocking}

For the sake of completeness, let us consider here the 
specialities of the situation
intermediate between the 
exponentially stable
`generic \phaselocking{}' 
taking place if $\Delta>0$
and {\chao} phase
evolutions
for which $\Delta<0$.
It is distinguished by the merging 
of the two eigenvalues of the matrix (\ref{eq25}).
Zero discriminant condition and Eq.\ (\ref{eq20a}) imply
$$
 \cosh\half P_0(T)
 \cos\half\phi_0(T)
-\half
 e^{\half P_0(T)} Q_0(T)\sin\half\phi_0(T)=
\mbox{\ either\ }1\mbox{\ or\ }-1
$$ 
and then it follows from (\ref{eq35x})
that the two-fold eigenvalue of $\mathbf{\Phi}$ which now is represented as follows
 \begin{eqnarray}
\lambda&=&
\half
e^{\half P_0(T)}
\left(
      (1+e^{-P_0(T)}) \cos\half\phi_0(T)
-Q_0(T)          \sin\half \phi_0(T)
\right)
 \end{eqnarray}
has also to be equal to either 1 or -1.

In the case $\Delta=0$ Eq.\ (\ref{eq:nfhtrse}) does not apply.
Instead, one may employ the 
following
explicit representation of the power of the $2\times2$ matrix
\begin{equation}
  \label{eq:tuyerr}
 \mathbf{\Phi}=\left(
\begin{array}{ll}
a&b\\c&d
\end{array}
\right)
\end{equation}
whose elements obey the constraint 
\begin{equation}
  \label{eq:rtuyerrr}
4b c+(a-d)^2=0
\end{equation}
(just meaning that the $\mathbf{\Phi}$ eigenvalues coincide)
times the 2-element column with arbitrary elements $A,B$:
\begin{eqnarray}
  \label{eq:jgyrtdf}
\mathbf{\Phi}^j
\left(
\begin{array}{l}
A\\B
\end{array}
\right)&=&
(\half(a+d))^j
\left[
{B\over 2c}
\left(
\begin{array}{c}
a-d\\2c
\end{array}
\right)
\right.
                       \nonumber\\
&&
\left.
+\left(
A-{a-d\over 2 c}B
\right)
\left(
\left(
\begin{array}{c}
1\\0
\end{array}
\right)
+{j\over a+d}
\left(
\begin{array}{c}
a-d\\2c
\end{array}
\right)
\right)
\right].
\end{eqnarray}
(It can be established, for example, by means of the mathematical induction.)
Since the eigenvalue does not vanish, 
$a+d\not=0$. Let us assume also, for a
while, that
$c\ne0\ne b$  which imply, in view of (\ref{eq:rtuyerrr}),
$a-d\ne0$.
Then the following analogue of Eq.\ (\ref{eq42}) arises:
\begin{equation}
  \label{eq:ftreh}
  C_j
=
{(a+d)C_0+((a-d)C_0+2b)j
\over
(a+d)+(2c C_0-(a-d))j
}
\end{equation}
It implies the existence of the limit
\begin{eqnarray}
  \label{eq:jftyr}
  \lim_{j\rightarrow\infty}C_j&=&
{(a-d)C_0+2b
\over
2c C_0-(a-d)
}=
2b{(a-d)C_0+2b
\over
4bc C_0-2b(a-d)
}
\nonumber\\
&=&
-2b
{(a-d)C_0+2b
\over
(a-d)((a-d) C_0+2b)
}
=
{-2b
\over
a-d
}
\nonumber\\
&=&
{a-d
\over
2c
}
\end{eqnarray}
which proves independent of $C_0$.
It also follows from 
(\ref{eq:ftreh}) that $C_j$ does not depend on $j$ (all the elements
of the sequence coincide) if and only if
\begin{equation}
  \label{eq:lkfhjgf}
  C_0=C_\infty\equiv{-2b\over a-d}={a-d\over 2c}.
\end{equation}
Thus, as opposed to the `generic \phaselocking{}' taking place for $\Delta>0$,
there is only a {\em single\/} initial phase which yields the `refined
steady' evolution. All the other phase evolutions converge, with 
the course of time $t$, to 
the latter which plays therefore the role of attractor.

Interestingly enough, 
the same phase function plays simultaneously the role of the
repeller  and 
all the distinct
phase functions approaching it are simultaneously `moving away'.
There is not contradiction here since the
phase functions live in fact on the closed circumference  and 
`the one side attracting' to a point is simultaneously `the another side
repelling'
from the same point.

To illustrate the simultaneous attraction/repelling property of a steady
phase function, let us 
compute the leading $j$-dependent contribution to  $C_j$.
The result is 
\begin{equation}
  \label{eq:rtyder}
  C_j=
C_\infty+{a+d\over 2c}j^{-1}+o(j^{-2})
\end{equation}
The terms shown do {\it not\/} depend on the initial phase (encoded
in $C_0$) which affects only the higher order contributions.
This means, in particular,
that all the phase functions approach their common steady limit
`from one side' (the leading contribution to the deviation from the
limit is common for all of them).
Then, obviously, the closer the initial phase is to the one
corresponding to the refined steady state
from {\em
  this\/} side, 
the less time is necessary for the reaching, in appropriate sense, the
steady limit.
On the other hand,
the `very long' phase evolution 
until it reaches some fixed vicinity of the ultimate steady state arises when 
the initial state is `very close' to the steady phase function from the
opposite, `wrong' side.
This behavior differs from the `generic' \phaselocking{} where 
the choices of the initial phases closed  to the phase of the stable steady evolution
always lead to the quick `monotonous' 
convergence to the limiting function irrespectively
to the initial relative angular direction.

This specialty can be inferred more rigorously from the consideration of
the analogue to Eq.\ (\ref{eq534}) which 
follows from (\ref{eq:ftreh}) and
now reads
\begin{eqnarray}
  \label{eq:jgudly}
  e^{\imi\phi(t'+T j)}&=&e^{\imi\phi_{(0)}(t')}\times
  \\
&&
{
2H G_+ (1+C_0\overline{{\cal F}_{(0)}(t')})+
(G_-+2H C_0)(2H-G_-\overline{{\cal F}_{(0)}(t')})j
\over
2H G_+ (1+C_0 {{\cal F}_{(0)}(t')})+
(G_-+2H C_0)(2H-G_- {{\cal F}_{(0)}(t')})j
},\nonumber
\end{eqnarray}
where
\begin{eqnarray}
  \label{eq:hftr}
  G_\pm&=&
(1\pm e^{-P_{(0)}(T)})\cos\half\phi_{(0)}(T) \mp  Q_{(0)}(T)\sin\half\phi_{(0)}(T),
\nonumber\\
H&=&
e^{-P_{(0)}(T)}\sin\half\phi_{(0)}(T)+ Q_{(0)}(T)\cos\half\phi_{(0)}(T)
\end{eqnarray}
(note that, in particular, $G_+=2e^{-\half P_{(0)}}\lambda$ equals either $2e^{-\half P_{(0)}}$
or $-2e^{-\half P_{(0)}}$ and does not vanish).
It determines the phase function for arbitrary $t$
expressing it
through the functions $\phi_{(0)},P_{(0)},Q_{(0)}$ specified on the
segment $[0,T]$.
The most substantial difference with Eqs.\ (\ref{eq:uyrvhyd}) and
(\ref{eq533}) is the 
{\em non-exponential\/} dependence on $j$. Now the convergence to the asymptotic 
phase function $\phi^{\infty}_{\bowtie}(t)$, 
defined $\mathrm{mod}\: 2\pi$ for any $t$ by the equation
\begin{equation}
  \label{eq:jryty}
e^
{\imi{\phi^{\infty}_{\bowtie}}(t)}=e^{\imi\phi_{(0)}(t')} 
\left.
{
2H-G_-\overline{{\cal F}_{(0)}(t')}
\over
2H-G_- {{\cal F}_{(0)}(t')}
}\right|_{t'=t-T[[t/T]]},
\end{equation}
 is linear in $j^{-1}\propto t^{-1}$.

Eq.\ (\ref{eq:jgudly}) can also be rewritten as follows
\begin{eqnarray}
  \label{eq:jgudlyr}
  e^{\imi\phi(t)-\imi\phi^{\infty}_{\bowtie}(t)}&=&
{
\overline{Z(t')}\delta C+
1 + 2H G_+^{-1} j\delta C
\over
Z(t')\delta C+
1 + 2H G_+^{-1}j\delta C
}, 
\\
\mbox{where}\;
Z(t')&=& 
{2 H {\cal F}(t')
\over
2 H - G_-{\cal F}(t')
}, \; \delta C=C_0+{G_-\over2H}.
\nonumber
\end{eqnarray}
For bounded $j$, the arbitrary constant $C_0$ can always be chosen making $\delta C$
so small that
the unity is
the dominating 
contribution in the both numerator and denominator 
in (\ref{eq:jgudlyr}). It makes
their ratio to
be as close to the unity as one desires. 
If further $j$ unboundedly increases, depending
on the
sign of $\delta C$,
the two distinct situations can occur.
Namely,
if $\delta C$ is of the same sign as  $HG_+^{-1}$, increasing $j$,
the ratio (\ref{eq:jgudlyr}) `monotonously' tend to the unity,
the less $\delta C$, the faster the limit is reached. This is the
case of the `true' situating of $C_0$ with respect to initial phase $C_\infty$ of
the steady phase function.
However, if $\delta C$ is still small but has the sign opposite to the sign
of $HG_+^{-1}$, increasing $j$, the  $j$-dependent contribution is subtracted
from the leading terms of the numerator and denominator (here the unity)
and the absolute value of their sum decreases reaching the minimum for
$j=[[|(H\delta C)^{-1}G_+|]]$. 
Simultaneously, the ratio  (\ref{eq:jgudlyr}) varies in some way 
and approaches the
initial closeness to the unity only for $j\simeq 2|(H\delta
C)^{-1}G_+|$ or greater. The further 
increasing of $j$ already leads to the 
`monotonous' converging to the unity with the rate $\sim j^{-1}$ as above.
However, the `convergence time' 
estimated as $2T|(H\delta C)^{-1}G_+|$ {\it increases\/}
as the deviation  $\delta C$ of the initial phase from the phase of
the steady function, having `wrong' sign, becomes smaller.

Finally, to abandon the temporal assumptions made above, it has to be noted that
the cases $c=0$ or $b=0$ implying $a=d$ which are not covered by the formulae above,
reveals qualitatively similar relationships. Their analysis
is carried out in a similar way, provided  
the following simple representations of 
$\mathbf\Phi^j$
$$
\left(
\begin{array}{ll}
a&0
\\
c&a
\end{array}
\right)^j=a^j
\left(
\begin{array}{ll}
1&0
\\
j c/a&1
\end{array}
\right),
\;\;
\left(
\begin{array}{ll}
a&b
\\
0&a
\end{array}
\right)^j=
a^j
\left(
\begin{array}{ll}
1&j b/a
\\
0&1
\end{array}
\right)
$$
are utilized. They
also lead to the convergence of the order $\sim j^{-1}$.

Resuming, in the case $\Delta=0$ not covered by the criterium
formulated in the form of the Theorem
\ref{gftdre}, the phase-locking understood as the 
asymptotic
convergence of all
phase functions to some steady one reproducing the own form
$\mathrm{mod}\:2\pi$ on each segment of $t$ variation of length $T$
takes place as well. However, it proves not exponential 
in time. Rather, the steady limit is approached as $\sim t^{-1}$.


\subsection*{Winding index}

Let us now derive yet another
important 
property of \phaselocking{} solutions of (\ref{eq1X}) following from
Eq.\ (\ref{eq534}).
It enables one to calculate the integer $k$
entering, in particular, equations (\ref{eq18}),(\ref{eq34}).
To that end,
calculating the logarithmic derivative of Eq.\ (\ref{eq:ogytgfdr})
and applying (\ref{eq85}),
one gets the equation
\begin{eqnarray}
\imi\dex\phi&=&\imi\dex\phi_{{\zeroth}}
+{C\dex \overline{\cal F}_{{\zeroth}}
\over
1+C\overline{\cal F}_{{\zeroth}}
}
-
{C\dex {\cal F}_{{\zeroth}}
\over
1+C{\cal F}_{{\zeroth}}
}
                        \nonumber\\
&=&\imi\dex\phi_{{\zeroth}}
+\imi C {e^{-\imi\phi_{{\zeroth}}}\dex t\over 1+C\overline{\cal F}_{{\zeroth}}}
+\imi C {e^{\imi\phi_{{\zeroth}}}\dex t\over 1+C{\cal F}_{{\zeroth}}}
                        \nonumber\\
&=&\imi\dex\phi_{{\zeroth}}
+2\imi C \Re\left[{e^{\imi\phi_{{\zeroth}}}\over 1+C{\cal F}_{{\zeroth}}}\right]\dex t
\end{eqnarray}
Integrating it on the interval $[0,T]$, one gets
\begin{equation}
\phi(T)-\phi(0)=\phi_{{\zeroth}}(T)+
2 C \Re\int^T_0
\left[{e^{\imi\phi_{{\zeroth}}}\over 1+C{\cal F}_{{\zeroth}}}\right]\dex t.
\nonumber
\end{equation}
Up to this point, the manipulations above do not go beyond Eq.\
(\ref{eq13}),
definitions and
identities.
The situation drastically changes if we substitute here in place of
arbitrary $C$ the constant  $C_\infty$ (\ref{eq539}) which, in the case $\Delta>0$
here assumed,
converts an arbitrary solution $\phi(t)$ of Eq.\ (\ref{eq1X}) to the
stable refined
\phaselocking{}
phase function
$\phi_\infty(t)$. Since the latter advances on each time step $[t,t+T]$
by
the strictly fixed increment
$\phi_\infty(t+T)-\phi_\infty(t)=2\pi k$, one gets
\begin{theo}
Let $\phi_{{\zeroth}}$ be the solution of (\ref{eq1X})
obeying the initial condition  $\phi_{{\zeroth}}(0)=0$ and
determining the 
complex valued function 
${\cal F}$ which, in turn, satisfies Eq.\ (\ref{eq85}) with initial condition
${\cal F}_{{\zeroth}}(0)=\imi$. Let also $\Delta[f]>0$, where
 $\Delta[f]$ is defined by Eq.\ (\ref{eq20a}). Then%
\footnote{There is a misprint in the corresponding equation in \cite{X7}.
}
\begin{equation}
k=
\frac{1}{2\pi}\phi_{{\zeroth}}(T)
-\frac{2}{\pi}\sin\half\phi_{{\zeroth}}(T)
\int^T_0
\Re\left[{e^{\imi\phi_{{\zeroth}}(t)}\over L_{max}
-2\sin\half\phi_{{\zeroth}}(T){\cal F}_{{\zeroth}}(t)}\right]
\dex t
                                    \label{eq861}
\end{equation}
is the {\em integer\/}
equal to the total sum (taking into account the orientation sign)
of the number of full revolutions
(the winding number) 
which any phase function,
except of $\phi_{\bowtie}$, 
will carry out
on the time steps of duration $T$
in its asymptotic steady \phaselocking{} state.
\end{theo}
{\it Remarks:}
\begin{itemize}
\item
The formula (\ref{eq861}) provides us with the substantiation of the 
statement made above which
concerns the constancy of the order of \phaselocking{} when the bias
parameters 
varies.
Indeed, in view of (\ref{eq861}) the order $k$
is continuous with respect to the
variables parametrizying the bias function $f$. In particular, 
$k$ continuously depends on $\iota_{\mbox{\scriptsize dc}}$
as far as $\Delta$ retains positive.
Hence it assumes a constant value on the 
$\iota_{\mbox{\scriptsize dc}}$
intervals over which $\Delta$ graph shown in Fig.\ \ref{f10} is
situated
above the horizontal coordinate axes (more generally, as the bias
function $f$ varies within the connected component of the
\phaselocking{} area).
\item
A more general representation of $k$ which operates with
not special but arbitrary solution of Eq.\ (\ref{eq1X}) 
can be derived from Eq.\ (\ref{eq861}), cf.\ the remark following Theorem 1.
However, as opposed to the case of criterion based on the
`universal' expression (\ref{eq:iroycd}), here the {\it two\/}
inequivalent formulae distinct by the opposite roles
of the left and right boundary points arise. 
Their difference is a complicated non-linear 
(and apparently non-trivial)
functional
which vanishes on all solutions of  (\ref{eq1X}),
provided the bias function corresponds to the \phaselocking{} phase evolution.
\end{itemize}

\section*{Conclusion}

The equation (\ref{eq1X}) and the properties of its solutions
seem to be of a considerable interest in view of several reasons.
First of all, this is, of course, 
their sound physical relevance following from
the extensive applications in 
the applied theory of electric activity of Josephson junctions
which employs Eq.\ (\ref{eq1X})
as the base of the efficient model 
of the junction phase dynamics \cite{X2,X3}
in the important
case of negligible role of junction
capacitance (overdamped junctions) \cite{X4,X5}.
On the other hand,  
Eq.\ (\ref{eq1X}) is of evident interest
in its own rights
from a pure mathematical point of view.
It provides us  
a remarkable example of apparently supreme simple
non-linear ODE which prove
associated with a linear problem%
\footnote{This point mentioned also in footnote \ref{ftr} (page \pageref{ftr})
is beyond the scope of present discussion.}
and, in view of such a link,
allows a deep exploration by analytic methods.
At the same time, in spite of its apparent simplicity,
it is definitely
far of being regarded as a mathematically trivial entity.
It suffices to say that the properties of Eq.\ (\ref{eq1X})
are still not completely understood even for sinusoidal bias
function $f=a+b\sin(\omega t+t_0)$, the case of a primary interest
from viewpoint of applications.

The relationships considered above does not exhaust 
the collection of rigorous
ones 
which (\ref{eq1X}) allows to establish 
by means of elementary technique.
However they are distinguished by the advantage
of universality being valid with fairly weak limitations 
on the class of allowable bias functions.

The introduction of the functional $\Delta[f]$ (\ref{eq20}) is the
central point of the approach.
Eq.\ (\ref{eq533}) is noteworthy as the explicit representation 
of the phase function for arbitrary $t$ through a single solution of
Eq.\ (\ref{eq1X})
computed on the finite interval $[0,T]$.
It is this equation which
allows to 
establish the convergence, 
in the case $\Delta[f]>0$, of any phase
function (except of
$\phi_{\bowtie}$, see Eq.\ (\ref{eq77a})) to an asymptotic limit
and to show that this limit coincides 
mod $2\pi$
with
$\phi_\infty $ defined by Eq.\ (\ref{eq77}). 
Thus Eq.\ (\ref{eq533}) yields the rigorous
model of the \phaselocking{} property allowing one to compute any of its
quantitative characteristic of interest.

In combination with Eq.\ (\ref{eq:uyrvhyd})
describing the long term phase evolution
in the opposite case
$\Delta[f]<0$,
the above relationships lead to the criterion
 of the asymptotic property of the \phaselocking{} (Theorem 1) %
which 
can be reformulated to 
operate with arbitrary single solution of Eq.\
(\ref{eq1X})
on a finite segment of the length $T$ and its derivates.

Another important result is the formula (\ref{eq861}) determining
the integer winding number $k$ (\phaselocking{} order)  through the same
easily computable data.
In physical terms, the latter integer quantity
is directly connected to the
average voltage applied across a junction in the \phaselocking{} state
which is the supported constant
and 
proves independent of the slow variations,
up to a certain extent, of the parameters, provided the period
$T$
is kept unchanged. This effect lies in the core of the
modern DC voltage standards \cite{X5}.

In view of the above remarks, 
the properties of  Eq.\ (\ref{eq1X}) and its solutions 
is
a fruitful area of the mathematical study which is worth
of a further development.

\end{document}